\documentclass[%
nofootinbib,
 amsmath,amssymb,
 aps,
floatfix,
10pt,
prd,twocolumn,
superscriptaddress]{revtex4-2}

\usepackage[dvipsnames]{xcolor}
\usepackage{graphicx}
\usepackage{dcolumn}
\usepackage{bm}
\usepackage[normalem]{ulem}
\usepackage{lipsum}

\usepackage[colorlinks]{hyperref}
\hypersetup{
  colorlinks=true,
  linkcolor={MidnightBlue},
  citecolor={OliveGreen},
  urlcolor={gray}
}

\newcommand{\appropto}{\mathrel{\vcenter{
  \offinterlineskip\halign{\hfil$##$\cr
    \propto\cr\noalign{\kern2pt}\sim\cr\noalign{\kern-2pt}}}}}

\begin{document}

\preprint{APS/123-QED}

\title{Chaos in Inhomogeneous Neutrino Fast Flavor Instability}

\author{Erick Urquilla}
\affiliation{Department of Physics and Astronomy, University of Tennessee, Knoxville, USA}
\affiliation{Escuela de Física, Facultad de Ciencias Naturales y Matemática, Universidad de El Salvador, Ciudad Universitaria, San Salvador, El Salvador}
\author{Sherwood Richers}%
\affiliation{Department of Physics and Astronomy, University of Tennessee, Knoxville, USA}


\begin{abstract}
    
    In dense neutrino gases, the neutrino-neutrino coherent forward scattering gives rise to a complex flavor oscillation phenomenon not fully incorporated in simulations of neutron star mergers (NSM) and core collapse supernovae (CCSNe). Moreover, it has been proposed to be chaotic, potentially limiting our ability to predict neutrino flavor transformations in simulations. To address this issue, we explore how small flavor perturbations evolve in the non-linear regime of the neutrino quantum kinetic equation within a narrow centimeter-scale region inside a NSM and a toy neutrino distribution. Our findings reveal that paths in the flavor state space of solutions with similar initial conditions diverge exponentially, exhibiting chaos. This inherent chaos makes the microscopic scales of neutrino flavor transformations unpredictable. However, the domain-averaged neutrino density matrix remains relatively stable, with chaos minimally affecting it. This particular property suggests that domain-averaged quantities remain reliable despite the exponential amplification of errors.

\end{abstract}

\maketitle

\section{Introduction}

    Neutrino flavor oscillation, proposed to explain the missing flux of neutrinos coming from the Sun, has ushered in a new era of neutrino physics beyond the standard model \cite{Kajita_1999,SNO2001} and a wide phenomenology of neutrino flavor transformation in astrophysics. Neutrinos propagate in a quantum superposition of weak interaction eigenstates that is modified by the forward interaction of neutrinos with charged leptons via the Mikheyev-Smirnov-Wolfenstein (MSW) effect \cite{Wolfenstein1978,MikheevSmirnov_1985,Mikheyev_and_Smirnov}. In addition, coherent neutrino-neutrino forward scattering produces a not well understood non-linear flavor oscillation in dense neutrino gases \cite{Pantaleone:1992eq,sigl1993general,Tamborra2021,duan2009,Capozzi_2022}.
    
    Neutrinos play a pivotal role in neutron star mergers (NSMs) (see \cite{radice2020dynamics,foucart2020brief,sarin2021evolution,shibata2019merger}) and CCSNe (see \cite{mezzacappa2020toward,Burrows_2021,janka2016physics,bethe1985revival}). These represent the sole locations in the universe, post the big bang, where neutrinos are generated at densities high enough to undergo strong interactions with each other. CCSNe arise when the core of a massive star collapses following the depletion of its nuclear fuel. The infalling material halts abruptly as the core attains nuclear density, giving rise to a bounce and a shock wave that propagates throughout the star. Neutrinos originating in the deep regions of the star drive the explosion, transporting energy from the collapsed core to the material falling below the shock front. In the delayed CCSNe scenario, this mechanism propels the stalled shock wave through the star, inducing the explosion \cite{mezzacappa2020physical,burrows2000supernova,mirizzi2016supernova,Janka_2012}. Numerous simulations have yielded predictions for CCSNe explosion energies, neutrino luminosities, electromagnetic signals, and ejected material \cite{Sandoval_2021,Kuroda_2022,burrows2019theoverarching,OConnor_2018,fiorillo_2023_sn_simulations_confront_1987a}. Several sources of uncertainty necessitate further investigation, including a consistent treatment of quantum neutrino transport \cite{Ehring2023fast,shalgar2022supernova}, the equation of state for matter beyond nuclear densities (e.g., \cite{lattimer_eos_review_2012,burgio_eos_review_2021}), and the nuclear reaction rates of heavy and highly unstable elements \cite{korobkin_2012_rprocess,Mumpower2015impact,Vassh_2019_fission,kullman_2023_nuclear_uncertainties}.
    
    NSMs and CCSNe are prime candidates for the production of much of the heavy nuclei of the universe (e.g., \cite{lattimer_schramm_1974,thielemann_2017_nsm_nucleosynthesis}). Neutrinos emitted from the resulting hot accretion disk and ejected material can significantly affect the production of neutrons and protons via the reactions
    \begin{eqnarray}
        p+\bar{\nu}_{e}&\leftrightarrow& n+e^{+}\\ 
        n+\nu_{e}&\leftrightarrow& p+e^{-}
    \end{eqnarray}
    that seed the conditions for slow and rapid neutron capture processes. Given that heavy lepton neutrinos do not partake in the creation and annihilation of protons and neutrons, a potential transformation of electron neutrinos into heavy flavors could significantly impact the ultimate production of heavy elements \cite{Radice2020thedynamics,curtis2023nucleosynthesis,wu2015effects,Lippuner_2017,Yoshida_2006}.
    
    NSMs are also important multimessenger events, as they emit gravitational waves, short gamma-ray bursts, kilonovae, electromagnetic afterglows and neutrinos \cite{burns2020neutron,fernandez2016electro,Margutii2021firstmul,radice_dynamics_2020,Radice_2018,Margalit_2019}. The gravitational wave event GW170817 \cite{Abbott2017gw170817} of two neutron stars colliding was widely studied in several multimessenger observations (e.g., \cite{Abbott2017multimessengerobs,Cowperthwaite_2017}). The observed kilonova light curve that resulted from radioactive decay of nuclei in the ejected material was the first direct detection of the synthesis of heavy elements through a rapid neutron capture process (r-process) \cite{metzger2017kilonovae,barnes2020physics}. From a theoretical and computational perspective, several general relativistic magnetohydrodynamics simulations have attempted to forecast the amount and composition of material ejected, the kilonova signal, neutrino luminosity, gravitational waves, and the final compact object resulting from the merger (e.g., \cite{Foucart2021estimating,Mosta_2020_amagnetar,Miller2019,Nedora_2022,Zhu2021fullygeneral,Metzger_2021,Li2021neutrinofast,Just2021neutrinoabs}). However, the neutrino flavor transformations have not yet been fully implemented in a NSM simulation.
    
    Despite its importance, the neutrino flavor transformation in dense neutrino environments is not fully understood. An ensemble of neutrinos that undergoes coherent forward scattering is an interacting quantum many-body system that could develop quantum entanglement. This can leave an imprint on the evolution histories of the neutrino flavor (see \cite{Patwardhan_2022} for a recent review). Quantum many-body systems exponentially increase complexity as the number of particles increases. Neglecting helicity and pair coherence the Hilbert space of an ensemble of $N$ interacting neutrinos and antineutrinos with $n_{f}$ number of flavors has a dimension of $n_{f}^{N}$. The exact solution for quantum many-body neutrino simulations has not involved more than $20$ neutrinos \cite{Patwardhan_2022,martin2023manybody,Ermal2020exactsolution,Pooja2023Entanglament}. Approximations of the quantum many body problem are based on compact representations of the wave function like tensor networks \cite{roggero2021dynamical,roggero2021entanglement,cervia2019entanglement} and other product state configurations \cite{paeckel2019timeevolu}. This allows hundreds of neutrinos to be simulated. Neglecting quantum entanglement, i.e., approximating expectation values of operator products as a product of the expectation values of individual operators, the neutrino and antineutrino ensemble can be described in the mean-field approximation by evolving $N$ pure quantum states as $n_{f}\times n_{f}$ density matrices. This reduces the Hilbert space to a dimension of $N n_F$. 
    
    The modeling approach for neutrino transport in mean-field approximation is termed neutrino quantum kinetics. This commonly applied approach enables the simulation of dense neutrino environments while inherently disregarding multi-particle quantum entanglement. Early many-body studies also support the validity of quantum kinetic theory for modeling dense astrophysical neutrino gases \cite{friedland2003neutrino, friedland2003many}. Recent studies have revealed discrepancies between simulations using the mean-field approximation and those employing many-body treatments, casting doubt on the validity of the former for modeling neutrinos in astrophysical environments. However, arguments have emerged suggesting that existing many-body calculations cannot be used to validate mean-field methods \cite{shalgar2023we, johns2023neutrino}, arguing that a many-body neutrino system with a small number of particles is expected to inherently differ from its mean-field counterpart. Unlike the mean-field approach, a self-consistent many-body framework should include momentum-changing incoherent scattering between neutrinos as part of the Hamiltonian. Since kinematic collision terms cannot reproduce many-body effects, these are expected to be absent in the mean-field approach. Furthermore, the limited number of neutrinos in many-body simulations complicates direct comparisons with environments suitable for mean-field approaches. In either case, even under the mean field approximation, a global implementation of the neutrino flavor transformation in NSMs and CCSNe is challenging due to the non-linear complexity of the flavor Hamiltonian and the small spatial and timescale of the flavor transformations.
    
    Although initial findings suggested that neutrino flavor oscillations were not crucial for the dynamics of CCSNe (e.g., \cite{Duan2011self-indiced,Dasgupta2012roleofcollective}), in the last decade, the growing understanding of fast flavor instabilities \cite{sawyer2005speed,sawyer2009multiangle, Sawyer_2016,richers2022fast,Capozzi_2022,volpe2023neutrinos} has ushered the possibility of fast neutrino flavor transformations on time and spatial scales of a few nanoseconds and centimeters in dense neutrino gases \cite{abbar_2018,Particle-in-cell,three_dimensions,Tamborra2021,Xinyu_2021,wu2021collective,xiong2021stationary,martin2021fast}. A prerequisite for the emergence of fast flavor instabilities is the existence of at least one direction in which equal numbers of neutrinos and antineutrinos are in motion. This condition manifests as a crossing in the angular distribution \cite{morinaga2022fast, dasgupta2022collective,abbar_2018,shalgar2019occurrence,nagakura2021and}. Conditions conducive to fast flavor conversion have been identified in the protoneutron star convection region of CCSNe \cite{Abbar2019Ontheoccurrence,glas2020fastneutrinof}, beneath the shock wave \cite{capozzi2021fastneutrinofla,Abbar2021Onthecharac,Nagaura2021wherewhenandwhy}, above the shock wave \cite{Bhattacharyya2021fastflavordep,Abbar_2022_supperssionoffast}, during the cooling phase of the protoneutron star \cite{Xiong_2020potentialimpact}, and in the NSM accretion disk \cite{grohs2022neutrino,george2020fastneutrinofla,Wu2017fastneutrinocon}.
    
    The non-linear evolution of flavor leads to intricate and seemingly chaotic dynamics, posing challenges for achieving convergence in dynamical calculations (e.g., \cite{richers_2022_code_comparison}). While chaos in the context of fast flavor instabilities remains unexplored, earlier investigations have hinted at chaos in bipolar oscillations within two-beam models (i.e., two momentum states) \cite{Hansen_2014,xiong2023symmetry}. Reference \cite{Hansen_2014} presents a method for obtaining the complete spectrum of twelve Lyapunov exponents and covariant Lyapunov vectors across various initial conditions with different symmetries that either maintain or disrupt the periodicity of bipolar oscillations. However, the long-term evolution of small perturbations and their maximum magnitudes were not explored. Reference \cite{xiong2023symmetry} demonstrates that small perturbations in neutrino flavor states yield a distinct flavor evolution history in a model incorporating fast axial symmetry-breaking modes. In such models, flavor oscillation modes either lack regularity or exhibit features resembling periodic behavior for a brief duration. The introduction of a small perturbation does not guarantee that the system will evolve in a manner where its evolution paths closely align.
    
    Future observations of CCSNe and NSMs neutrinos will provide crucial scientific insights into the collapsing stellar core (e.g., \cite{Horiuchi_2018}) and nucleosynthesis conditions. The exponential amplification of errors in simulations, due to chaos, may limit our ability to predict the evolution of neutrino flavor within CCSNe and NSMs, hindering our capacity to compare theoretical predictions with observational data. To shed light on the problem, we extend the study of chaos to a more realistic distribution of neutrinos that exhibit fast flavor instabilities. We conduct two simulations of dense neutrino gases with distinct angular distributions, one of which involves a narrow domain a few centimeters wide, capturing a snapshot of a NSM. The resolution scale is defined by a domain of $1\times1\times 64$ cm divided into $1\times 1\times 1024$ cells, each accommodating $24,088$ momentum states. This setup allows for flavor anisotropies and inhomogeneities. The primary goal is to characterize the evolution of small perturbations in neutrino flavor states over time, specifically quantifying the sensitivity of the results to the initial conditions. This endeavor will shed light on how chaos affects our capability to predict the evolution of neutrino flavor in dense astrophysical neutrino gases within simulations.
    
    In Section~\ref{sec:background:chaos}, we briefly overview the standard theory of chaos in dynamical systems. Neutrino quantum kinetics is reviewed in Section~\ref{sec:quantum_kinetics}, while Section~\ref{sec:methods} outlines our numerical approach to simulating neutrino flavor transformation. Evidence of chaos is presented in two extensive one-dimensional simulations in Section~\ref{sec:results}. We further explore the unpredictable nature of chaos and its impact on both large and small scales of neutrino flavor transformation in the same Section~\ref{sec:results}. Concluding remarks are provided in Section~\ref{sec:conclusions}.

\section{Chaos}
\label{sec:background:chaos}
    
    The discovery by E. Lorenz of an atmospheric convection model exhibiting irregular and unpredictable behavior formally marked the beginning of the study of chaotic systems \cite{lorenz_1963}. Chaotic systems are non-linear dynamical systems characterized by erratic and irregular complex behavior. Although these systems are fundamentally deterministic — meaning precise knowledge of initial conditions allows for the prediction of future behavior — even a minute variation in these conditions, such as measurement uncertainty, results in an entirely different prediction. This contrasts with non-chaotic systems, where the approximate present determines the future approximately. In chaotic systems, the approximate present leads to an entirely different future (see \cite{baker_gollub_1996,taylor2005classical, strogatz2014nonlinear,zaslavsky2005hamiltonian,hilborn2000chaos} for modern expositions of chaotic systems).
    
    Since Lorenz's findings, many dynamical systems have been discovered to exhibit chaos, revealing how chaos and complexity can emerge from seemingly simple dynamic systems such as the logistic equation \cite{may1974biological}, a driven damped pendulum \cite{huberman1980noise}, or a double pendulum \cite{shinbrot1992chaos}. Chaos, rather than being an isolated behavior of specific dynamical systems, is a universal property of complexity in all of them \cite{hilborn2000chaos}.
    
    Consider a general non-linear dynamical system characterized by a state space of dimension $n+1$ with coordinates $r_i$, which follows the governing equation
    \begin{eqnarray}
        \dot{\vec{r}}=\vec{F}(r_i).
        \label{dynamicalsystem}
    \end{eqnarray}
    The solution $\vec{r}_{t}$ with initial conditions $\vec{r}_{t_0}$ traces a path that we assume confined to a finite volume in the state space, i.e. $r_i \in [a_i,b_i]$.
    
\begin{figure}[!htbp]
    \includegraphics[width=0.85\linewidth]{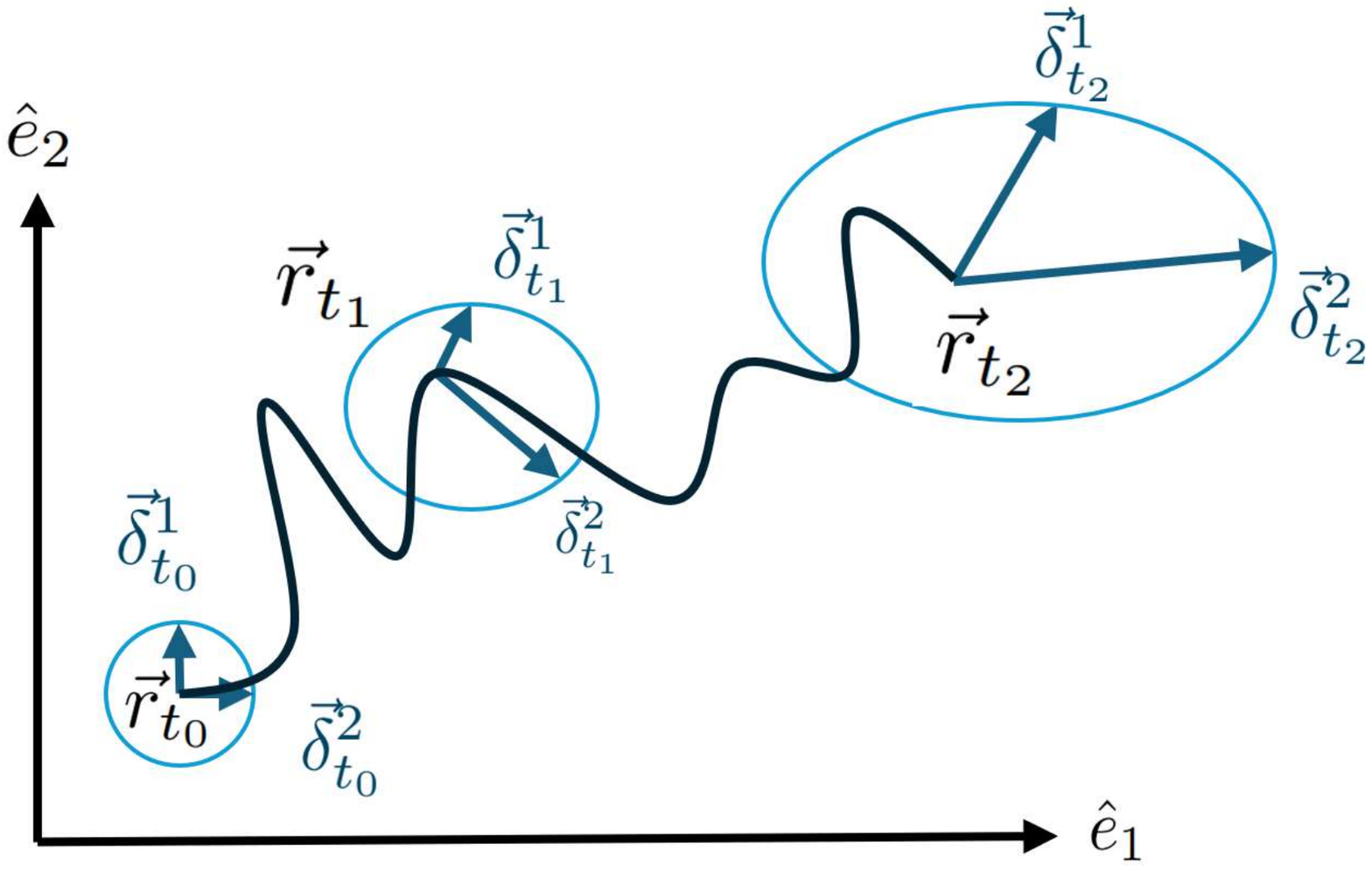}
    \caption{\label{fig: Lyapunov_figure} 
        Graphical representation of a chaotic dynamical system governed by equation \eqref{dynamicalsystem}. Consider a $1$-sphere of initial conditions $\vec{\delta}_{t_0}^{k}$ around $\vec{r}_{t_0}$, where $k=1,\,2$. As time progresses, this $1$-sphere morphs into a $1$-ellipsoid, as demonstrated at $t_{1}$ and $t_{2}$, with each perturbation ($\vec{\delta}_{t}^{1}$ and $\vec{\delta}_{t}^{2}$) experiencing variable expansion and contraction. In a chaotic dynamical system, the transformation of the $1$-ellipsoid's form is governed by the equation \eqref{eq: exponential_trend_of_perturbations} and at least one Lyapunov exponent ($\lambda_{k}$) is positive. The exponential divergence of two nearly identical states implies that systems with minor, barely perceptible differences in their initial conditions will begin to exhibit divergent behavior in the near future, significantly diminishing the capacity for accurate prediction.}
    \end{figure}

    A frequently used strategy to identify chaos involves quantifying the sensitivity of the outcomes relative to the initial conditions. One should contemplate an $n$-sphere of initial conditions $\vec{r}_{t_0}+\vec{\delta}_{t_0}^{k}$, centralized at $\vec{r}_{t_0}$ (refer to a two-dimensional depiction at $t_{0}$ in figure \ref{fig: Lyapunov_figure}). Here, $\vec{\delta}_{t_0}^{k}$ is a small perturbation and $k=1, \,2, \,...\, n+1$ symbolizes the state space dimensionality of the dynamical system. As time progresses, the dynamical system's non-linearity transforms the $n$-sphere into an $n$-ellipsoid (observe the one-ellipsoids at $t_{1}$ and $t_{2}$ in Figure \ref{fig: Lyapunov_figure}), with each dimension experiencing varying rates of expansion and contraction.  If the shape of the $n$-ellipsoid evolves according to
    \begin{eqnarray}
        |\,\vec{\delta}_{t}^{k}\,|\approx |\,\vec{\delta}_{t_0}^{k}\,|\, e^{\lambda_{k}t}, \label{eq: exponential_trend_of_perturbations}
    \end{eqnarray}
    $\lambda_{k}$ is a Lyapunov exponent of the dynamical system. A positive Lyapunov exponent indicates instability, while a negative one indicates stability. Neutral stable regions are implied by a zero Lyapunov exponent. A non-linear dynamical system is chaotic if it possesses at least one positive Lyapunov exponent. In chaotic systems, the Lyapunov exponent characterizes the average rate of exponential divergence of nearby trajectories in state space \cite{wolf1985determining}. If a perturbation evolves for an extended period, the largest Lyapunov exponent will dominate the shape of the $n$-sphere, and the perturbations will follow the direction of maximum divergence \cite{baker_gollub_1996,hilborn2000chaos}. The exponential divergence of two nearly identical states suggests that systems with subtle, hard-to-notice distinctions in the initial conditions will soon start to behave differently, and the ability to make forecasts will be rapidly reduced. The magnitude of the Lyapunov exponent indicates the rate at which the system's behavior becomes unpredictable.
    
    In contrast to linear stability analysis, where the governing equations can be simplified to a linear form with the aim of identifying the existence of growing exponential modes at a specific point in the state space, the analysis of chaos investigates the evolution of perturbations in the non-linear regime of the governing equations. Non-linearity is a requirement for the appearance of chaos, as is the confinement of the solution trajectories to a finite volume in the state space $r_i \in [a_i,b_i]$. As demonstrated in Appendix \ref{appendixb}, solutions of the neutrino quantum kinetic formalism fall into this category when collisions are not present.
    
    We should also note that, given that the solutions of the dynamical system are confined to a certain volume in the state space, the perturbation cannot expand exponentially beyond the size of the state space that limits the solutions. Equation \eqref{eq: exponential_trend_of_perturbations} is valid in this time domain. This constraint, together with the exponential separation of the solutions, implies that the solutions are continuously flexing and unfurling over different regions of the state space, erasing the intricate structure of the solutions and producing new information. The Lyapunov exponent measures the rate at which the system sheds information over time, usually expressed in bits per second \cite{wolf1985determining}.
    
\section{Neutrino quantum kinetics}
\label{sec:quantum_kinetics}
    
    A proper treatment of a flavor neutrino field in a dense environment involves a seven-dimensional distribution function for neutrinos $f_{ab}(t,\textbf{x},\textbf{p})$ and antineutrinos $\bar{f}_{ab}(t,\textbf{x},\textbf{p})$ ($n_{F}\times n_{F}$ matrices, where $n_{F}$ is the number of neutrino and antineutrino flavors) whose elements are the expectation values of bilinear creation and annihilation operators $< a_{a}^{\dagger}a_{b}>$ and $< b_{a}^{\dagger}b_{b}>$ in the mean field approach. The diagonal terms ($a=b$) of $f_{ab}(t,\textbf{x},\textbf{p})$ and $\bar{f}_{ab}(t,\textbf{x},\textbf{p})$ represent the neutrino and antineutrino occupation numbers for each flavor ($a,\,b \in \{e,\, \mu,\,\tau\} $), and the off-diagonal term ($a \neq b$) represents the flavor correlation \cite{Tamborra2021,Capozzi_2022}. The time evolution of the neutrino distribution function is governed by the quantum kinetic equation (QKE)
    \begin{equation}
            (\partial_t+\textbf{v}\cdot \nabla_{x}) f_{ab}=C_{ab}-\frac{i}{\hbar}\left[H,f\right]_{ab}.
            \label{eq: QKE}
    \end{equation}
    Barred quantities ($\bar{f}_{ab}$, $\bar{C}_{ab}$ and $\bar{H}$) need to be substituted to arrive at the equation for antineutrino distributions. The matrix $C_{ab}$($\bar{C}_{ab}$) is the collision term which accounts for the non-forward scattering of neutrinos(antineutrinos). The first term on the left accounts for the explicit time dependence of the distribution function and the second term accounts for the drift that is caused by the particles that are streaming freely. We ignore other external forces like gravity and hypothetical electromagnetic interactions. The Hamiltonian $H_{ab}$ accounts for the flavor transformation due to mixing between mass and flavor eigenstates $H^{\mathrm{vacuum}}_{ab}$, the coherent forward scattering of neutrinos with background matter (other than neutrinos) $H^{\mathrm{matter}}_{ab}$, and the coherent forward scattering of neutrinos with others neutrinos and antineutrinos $H^{\mathrm{\nu-\nu}}_{ab}$. These are combined as
    \begin{equation}
        H_{ab}=H^{\mathrm{vacuum}}_{ab}+H^{\mathrm{matter}}_{ab}+H^{\mathrm{\nu-\nu}}_{ab},
    \end{equation}
    where
    \begin{eqnarray}
        H^{\mathrm{vacuum}}_{ab}&=& U_{ac}\left[\sqrt{\textbf{p}^2 c^2+ m_c ^2 c^4 }\delta_{cd}\right]U_{db}^{\dagger} \\
        H^{\mathrm{matter}}_{ab}&=& \sqrt{2} G_F (\hbar c)^{3} \left[\left(n_a-\bar{n}_{a}\right)-\hat{\textbf{p}}\cdot \left( \textbf{f}_a -\bar{\textbf{f}}_a\right) \right] \delta_{ab} \\    
        H^{\mathrm{\nu-\nu}}_{ab}&=& \sqrt{2} G_F (\hbar c)^{3} \left[\left(n_{ab}-\bar{n}^*_{ab}\right)-\hat{\textbf{p}}\cdot\left( \textbf{f}_{ab} -\bar{\textbf{f}}_{ab}^* \right) \right].
    \end{eqnarray}
    Repeated indices are summed. $U$ is the Pontecorvo-Maki-Nakagawa-Sakata (PMNS) unitary mixing matrix that for $n_{F}=3$ takes the form 
    \begin{widetext}
    \begin{eqnarray}
        U=\begin{pmatrix}
        c_{12}c_{13} & s_{12}c_{13} & s_{13}e^{-i\delta_{cp}}\\
        -s_{12}c_{23}-c_{12}s_{13}s_{23}e^{i\delta_{cp}} & c_{12}c_{23}-s_{12}s_{13}s_{23}e^{i\delta_{cp}} & c_{12}s_{23}\\
        s_{12}s_{23}-c_{12}s_{13}s_{23}e^{i\delta_{cp}} & c_{12}s_{23}+s_{12}s_{13}c_{23}e^{i\delta_{cp}} & c_{12}c_{23}
        \end{pmatrix},
    \end{eqnarray}
    \end{widetext}
    where $s_{ij}=\sin \theta_{ij}$ and $c_{ij}=\cos \theta_{ij}$. $\theta_{ij}$ are the neutrino mixing angles, $\delta_{cp}$ is the charge-parity violation phase and $m_{c}$ is the neutrino mass in the masses eigenstates for $c\in\{1,\,2,\,3\}$. In this work we consider $\theta_{12}=33.82^{\circ}$, $\theta_{13}=8.61^{\circ}$, $\theta_{23}=48.33^{\circ}$, $\delta_{cp}=222^{\circ}$, $\delta m_{21}^{2}=7.39\times 10^{-5}\,\mathrm{eV}^{2}$ and $\delta m_{32}^{2}=2.449\times 10^{-3}\,\mathrm{eV}^{2}$ under the neutrino normal mass ordering as in reference \cite{particle2020review}, although the calculations are in a regime where these terms are of negligible import. $\textbf{p}$ is the neutrino momentum. $n_{a}$($\bar{n}_{a}$) denote the scalar number density and $\textbf{f}_{a}$($\bar{\textbf{f}}_{a}$) denote the number flux density for leptons(anti-leptons) of flavor $a$. The Hermitian neutrino number density $n_{ab}$ and number density flux $\textbf{f}_{ab}$ matrices come from the neutrino distribution function as follows:
    \begin{eqnarray}
        n_{ab}&=&\frac{1}{(\hbar c)^{3}}\int d\hat{\textbf{p}}\int \frac{\text{p}^2d\text{p}}{(2\pi)^{3}} f_{ab}, \\
        \textbf{f}_{ab}&=&\frac{1}{(\hbar c)^{3}}\int d\hat{\textbf{p}}\int \frac{\text{p}^2d\text{p}}{(2\pi)^{3}} f_{ab} \hat{\textbf{p}}.
    \end{eqnarray}
    The antineutrino equations are given by substituting barred quantities as 
    \begin{eqnarray}
        \bar{H}^{\mathrm{vacuum}}_{ab}&=&{H^{\mathrm{vacuum}}_{ab}}\\
        \bar{H}^{\mathrm{matter}}_{ab}&=&-{H^{\mathrm{matter}}_{ab}}^*\\
        \bar{H}^{\mathrm{\bar{\nu}-\bar{\nu}}}_{ab}&=&-{H^{\mathrm{\nu-\nu}}_{ab}}^*.
    \end{eqnarray}
    In this work we suppress the matter term in the Hamiltonian to focus on the flavor transformation produced by the neutrino-neutrino coherent forward scattering mechanism. 
    
\section{Methods}
\label{sec:methods}

    We employ mean-field neutrino quantum kinetics, taking into account two angular degrees of freedom and allowing for spatial flavor inhomogeneities in one dimension, while preserving homogeneity in the other two dimensions. We carry out simulations of mono energetic neutrinos. We make use of the multidimensional neutrino quantum kinetics code Emu, which we describe in Section~\ref{sec:emu}.  The pair of initial conditions for the neutrino systems under study is detailed in Section~\ref{sec:initial_conditions}.
    
    \subsection{Emu}
    \label{sec:emu}
        
        Emu \cite{Particle-in-cell} is a particle-in-cell code designed for simulating neutrino flavor transformations. It solves the QKE without considering collisions under the mean-field approximation. Emu utilizes computational particles to discretize the neutrino field into distinct packages and tracks the evolution of each computational particle's position on a background grid. This approach efficiently accounts for the advection term in the Liouville operator $\textbf{v}\cdot \nabla_{x}$ by moving the computational particles in accordance with the equation
        \begin{eqnarray}
            \frac{\partial \textbf{r}}{\partial t}&=&c\hat{\textbf{p}}\\
            \frac{\partial \textbf{p}}{\partial t}&=&0
        \end{eqnarray}
        where $\mathbf{r}$ is the position vector of the particle and $\mathbf{p}$ is the momentum vector of the particle. This implies that there is no momentum exchange due to incoherent scattering, creation and annihilation of neutrinos and antineutrinos and energy exchange $C_{ab}=\bar{C}_{ab}=0$ so that each computational particle in EMU satisfies
        \begin{eqnarray}
            \frac{\partial N}{\partial t}&=&\frac{\partial \bar{N}}{\partial t}=0\\
            \frac{\partial E}{\partial t}&=&0,
        \end{eqnarray}
        where $N$ and $\bar{N}$ are the number of physical neutrinos and antineutrinos that a computational particle represents.
        
        Each computational particle carries two quantum states defined by the Hermitian density matrices $\rho$ and $\bar{\rho}$, representing the physical neutrinos and antineutrinos carried by the particles. The flavor quantum state evolves in time according to the Schrödinger equation
        \begin{eqnarray}
            \frac{\partial \rho}{\partial t}&=&-\frac{i}{\hbar}\left[H,\rho\right]\\
            \frac{\partial \bar{\rho}}{\partial t}&=&-\frac{i}{\hbar}\left[\bar{H},\bar{\rho}\right].
        \end{eqnarray}
        
        To compute the Hamiltonian based on a discrete approximation of the neutrino field, EMU employs a deposition and interpolation algorithm to estimate the number density of neutrinos and antineutrinos ($n_{ab}$ and $\bar{n}_{ab}$) and the number density flux ($\textbf{f}_{ab}$ and $\bar{\textbf{f}}_{ab}$) at every position of the particle. During each simulation time step, EMU models each computational particle with an extended shape centered on the particle position, with an extent comparable to the size of the grid cell. This is used to combine the quantum states of the computational particles that overlap over a single cell based on the parameterization of a second-order shape function. This combination attributes a distribution of neutrinos and antineutrinos from each particle to each cell of the grid. Subsequently, the neutrino distribution moments are interpolated from the grid cell distribution to the location of each computational particle, using the same second-order shape function. The quantum state is then integrated using a fourth-order Runge-Kutta method in an efficient and scalable manner based on the AMReX framework, ensuring performance portability to both CPU and GPU hardware.
        
    \subsection{Initial conditions}
    \label{sec:initial_conditions}
        
        We study the presence of chaos in two separate simulations. In the first (Section~\ref{Initial conditions: Fiducial simulation}) we look at a well-understood toy model, and in the second (Section~\ref{Initial conditions NSM snapshot simulation}) we extract the neutrino distribution from a multidimensional NSM simulation. In both cases, we construct a simulation domain that is one-dimensional in space (i.e., assuming homogeneity in the $x$ and $y$ directions) using a one-dimensional array of $1\times 1\times 1024$ cells embedded in a domain of $1\times1\times64$ cm with periodic boundary conditions. The final flavor content resulting from the flavor instabilities in one-dimensional simulations provides a reasonably accurate approximation compared to computationally expensive three-dimensional simulations \cite{three_dimensions}. We simulate $24,088$ computational particles initialized at the center of each cell, with isotropically distributed directions ensuring that each computational particle represents the same solid angle. This significantly larger particle count per cell is chosen as we observe that the properties of chaos are notably more sensitive to numerical errors than the global distribution averages investigated in previous works \cite{Particle-in-cell,three_dimensions}.

    \subsubsection{Fiducial simulation}
    \label{Initial conditions: Fiducial simulation}
        
        \begin{figure}[!htbp]
            \centering
            \includegraphics[width=0.8\linewidth]{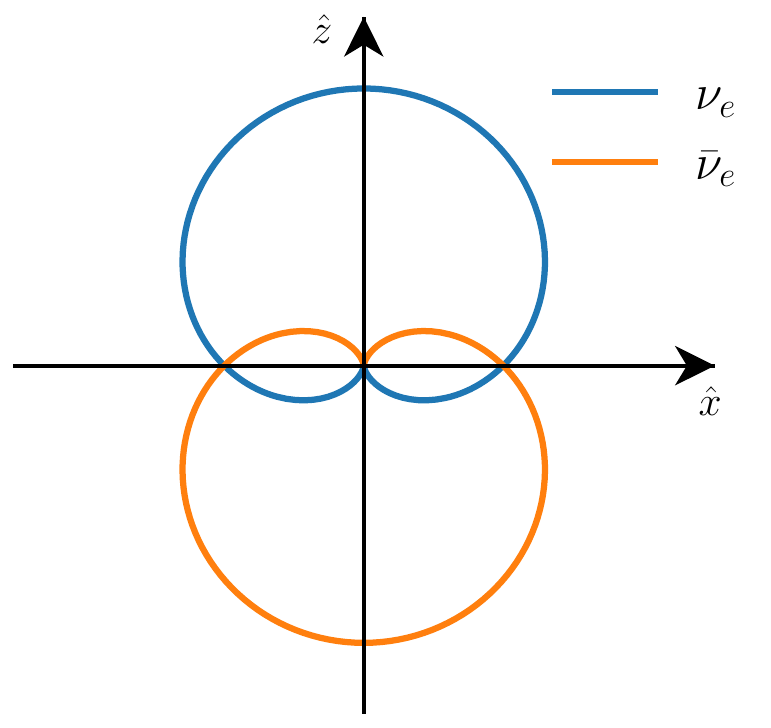}
            \caption{
                \label{fig:fiducial_angular_distribution} Initial angular distribution for neutrinos and antineutrinos in the one-dimensional Fiducial simulation. A rotation of the curves around the $\hat{z}$ direction ($\hat{z}$ axial symmetry) reproduces the three-dimensional neutrino angular distribution. The neutrino and antineutrino number density emitted per solid angle is proportional to the distance of the curves from the origin.
            }
        \end{figure}
        
        In this simulation, the neutrino and antineutrino distributions are purely electron flavor. The total neutrino and antineutrino fluxes point in opposite directions with a flux factor ($\textbf{f}_{ab}/n_{ab}$ and $\bar{\textbf{f}}_{ab}/\bar{n}_{ab}$) of $1/3$ and a number density of $4.89 \times 10^{32}$ cm$^{-3}$. Although these parameters are purely a toy problem, similar total neutrino number densities occur in core-collapse supernovae at a radius of approximately $80$ km between $200$ and $300$ ms post-bouncing \cite{abbar2019occurrence,glas2020fast} and at a radius of $40$ km in neutron star merger accretion disks \cite{wu2017imprints}. In this problem, the initial neutrino angular distributions satisfy
        \begin{eqnarray}
            \frac{dn_{\nu_e}}{d\bf{\Omega}}&=&\frac{n_{\nu_e}}{4\pi}(1+\cos\theta)\\
            \frac{dn_{\bar{\nu}_e}}{d\bf{\Omega}}&=&\frac{n_{\bar{\nu}_e}}{4\pi}(1-\cos\theta)\\
            \frac{dn_{\nu_\mu}}{d\bf{\Omega}}&=&\frac{dn_{\nu_\tau}}{d\bf{\Omega}}=\frac{dn_{\nu_\tau}}{d\bf{\Omega}}=\frac{dn_{\bar{\nu}_\mu}}{d\bf{\Omega}}=\frac{dn_{\bar{\nu}_\tau}}{d\bf{\Omega}}=0
        \end{eqnarray}
        where $n_{\nu_\alpha}$($n_{\bar{\nu}_\alpha}$) is the neutrino(antineutrino) number density of flavor $\alpha$. $d\bf{\Omega}$ is a differential solid angle in momentum space centered on the momentum direction $\hat{\textbf{p}}$. The direction $z$ defines the direction from which $\theta$ is defined zero. Figure \ref{fig:fiducial_angular_distribution} shows the initial angular distribution in the $xz$ plane. The linear angular distribution exhibits a strong lepton-number crossing in the $xy$ plane. Previous work shows that this distribution exhibit fast flavor instabilities that match the unstable modes predicted by linear stability analysis \cite{Particle-in-cell}. The input parameters of this simulation are publicly available in the EMU code \footnote{\texttt{https://github.com/AMReX-Astro/Emu}}. 
        
        To seed the instability and set a small initial amplitude for fast unstable flavor modes, the quantum state of each computational particle is initialized in nearly the pure electron flavor state, with a small random perturbation on the non-diagonal electron-muon and electron-tau components:
        \begin{eqnarray}
            \rho = 
            \begin{bmatrix}
                1-\epsilon_{\mu}-\epsilon_{\tau} & \alpha\left(Q+Qi\right) & \alpha\left(Q+Qi\right) \\
                \rho^{*}_{e\mu} & \epsilon_{\mu} & 0 \\
                \rho^{*}_{e\tau} & 0 & \epsilon_{\tau} 
            \end{bmatrix},
        \end{eqnarray}
        \begin{eqnarray}
            \bar{\rho} = 
            \begin{bmatrix}
                1-\epsilon_{\mu}-\epsilon_{\tau} & \alpha\left(Q+Qi\right) & \alpha\left(Q+Qi\right) \\
                \bar{\rho}^{*}_{e\mu} & \epsilon_{\mu} & 0 \\
                \bar{\rho}^{*}_{e\tau} & 0 & \epsilon_{\tau} 
            \end{bmatrix}.
        \end{eqnarray}
        Here, $Q$ is a random number generated between $-1$ and $1$ each time it appears, and $\alpha$ is the strength of the random perturbation, which we set as $10^{-6}$. $\epsilon_{\mu}$ and $\epsilon_{\tau}$ are computed once the random non-diagonal components are generated in order to satisfy the conservation of the length of the flavor polarization vector and the unit trace of the density matrix.
        
    \subsubsection{NSM snapshot simulation}
    \label{Initial conditions NSM snapshot simulation}

        \begin{table*}
        
        \caption{\label{tab:table2} Initial neutrino and antineutrino total number densities and fluxes for the NSM snapshot simulation.}
        
            \newlength\q
        
            \setlength\q{\dimexpr .5\textwidth -2\tabcolsep}
        
            \noindent\begin{tabular}{p{\q}p{\q}}
        
            \hline 
            \hline 
        
            Number density $n_{ab}$ ($\mathrm{cm}^{-3}$) & Flux factor $\textbf{f}_{ab}/n_{ab}$ \\ 
        
            \hline 
        
            $n_{ee}=1.422\times 10^{33}$ & $\textbf{f}_{ee}/n_{ee}=0.0974\hat{\boldsymbol{e}}_{x}+0.0421\hat{\boldsymbol{e}}_{y}-0.1343\hat{\boldsymbol{e}}_{z}$ \\
            
            $n_{\mu\mu}=4.913\times 10^{32}$ & $\textbf{f}_{\mu\mu}/n_{\mu\mu}=-0.0216\hat{\boldsymbol{e}}_{x}+0.0743\hat{\boldsymbol{e}}_{y}-0.5354\hat{\boldsymbol{e}}_{z}$\\
            
            $n_{\tau\tau}=4.913\times 10^{32}$ & $\textbf{f}_{\tau\tau}/n_{\tau\tau}=-0.0216\hat{\boldsymbol{e}}_{x}+0.0743\hat{\boldsymbol{e}}_{y}-0.5354\hat{\boldsymbol{e}}_{z}$\\
            
            $\bar{n}_{ee}=1.915\times 10^{33}$ & $\bar{\textbf{f}}_{ee}/n_{ee}=0.0723\hat{\boldsymbol{e}}_{x}+0.0313\hat{\boldsymbol{e}}_{y}-0.3446\hat{\boldsymbol{e}}_{z}$\\
            
            $\bar{n}_{\mu\mu}=4.913\times 10^{32}$ & $\bar{\textbf{f}}_{\mu\mu}/n_{\mu\mu}=-0.0216\hat{\boldsymbol{e}}_{x}+0.0743\hat{\boldsymbol{e}}_{y}-0.5354\hat{\boldsymbol{e}}_{z}$\\
            
            $\bar{n}_{\tau\tau}=4.913\times 10^{32}$ & $\bar{\textbf{f}}_{\tau\tau}/n_{\tau\tau}=-0.0216\hat{\boldsymbol{e}}_{x}+0.0743\hat{\boldsymbol{e}}_{y}-0.5354\hat{\boldsymbol{e}}_{z}$\\
            
            \hline 
            \hline 
            
        \end{tabular}
        \end{table*}
        
        We evolve the quantum flavor states of neutrinos in a NSM snapshot after $5$ ms post-merger from the classical global general relativistic two-moment radiation hydrodynamics simulation in \cite{foucart2016impact}, at a location approximately $40^\circ$ from the accretion disk plane and at $30$ km from the compact object center as in \cite{grohs2022neutrino}. The total number densities and fluxes of neutrinos and antineutrinos is shown in Table \ref{tab:table2}.
        
        To maintain consistency with the assumptions made in the original NSM simulation, each computational particle carries a number of neutrinos and antineutrinos that satisfy the angular distribution determined by the maximum entropy closure \cite{1994ApJ...433..250C}. This maximizes the angular entropy of the energy-integrated distribution while conforming to the total neutrino and antineutrino number densities and fluxes outlined in Table \ref{tab:table2}. Specifically, the initial distribution function is given by
        \begin{eqnarray}
            f_{aa}(\vec{x},\vec{p},t) = \frac{n_{aa}}{2 \pi}\frac{Z}{\sinh(Z)} e^{Z\cdot \cos \theta},
        \end{eqnarray}
        where $n_{aa}$ is the number density of flavor $a$, and $\theta$ is the angle between the momentum direction and the flux direction. Barred quantities ($\bar{f}_{aa}$ and $\bar{n}_{aa}$) need to be substituted for antineutrino distributions. The factor $Z$ is numerically calculated to yield the expected total number density flux outlined in Table \ref{tab:table2}. This distribution exhibits a angular directions of equal fluxes of neutrinos and antineutrinos \cite{grohs2022neutrino}, indicating its unstable to fast flavor transformations (e.g., \cite{morinaga2022fast}).

\section{Results}
\label{sec:results}

    We explore how small perturbations in the flavor state of neutrinos and antineutrinos evolve over time using Lyapunov exponents (see Section \ref{sec:background:chaos}), on both macroscopic and microscopic levels of the neutrino flavor transformation. The macroscopic scales are represented by the domain-averaged density matrix, while the microscopic scales are computational particles that represent the smallest spatial resolution of the neutrino distributions in the simulations.
    
    We first simulate the neutrino flavor transformation in a domain a few centimeters wide, located above the accretion disk $5$ ms after the merger of two neutron stars \cite{foucart2016impact} (see Section \ref{Initial conditions NSM snapshot simulation} for a description of the simulation setup). This is a noteworthy area for the generation of r-process nuclei \cite{metzger2017kilonovae,barnes2020physics} and the neutrinos are unstable to the fast flavor instability, making this region an optimal testing ground for researching chaotic properties in neutrino flavor evolution. We also simulate a well-studied artificial neutrino distribution in which one third of the total number of neutrinos and antineutrinos travel in opposite directions (see Section \ref{Initial conditions: Fiducial simulation} for a description of the simulation setup). This is an extreme and unique initial condition that is useful for understanding of the patterns of chaos. Although, the three-dimensional neutrino flavor transformation for these systems have been reported in \cite{three_dimensions,grohs2022neutrino}, we conduct high-resolution one-dimensional simulations to precisely compute the Lyapunov exponents.
    
    Essential data have been archived and are publicly accessible at \cite{urquilla_2023_10447839}, with additional data available upon request.

    \subsection{Overall Dynamics}
    \label{Overall Dynamics}
        
        \begin{figure}[!htbp]
            \includegraphics[width=0.95\linewidth]{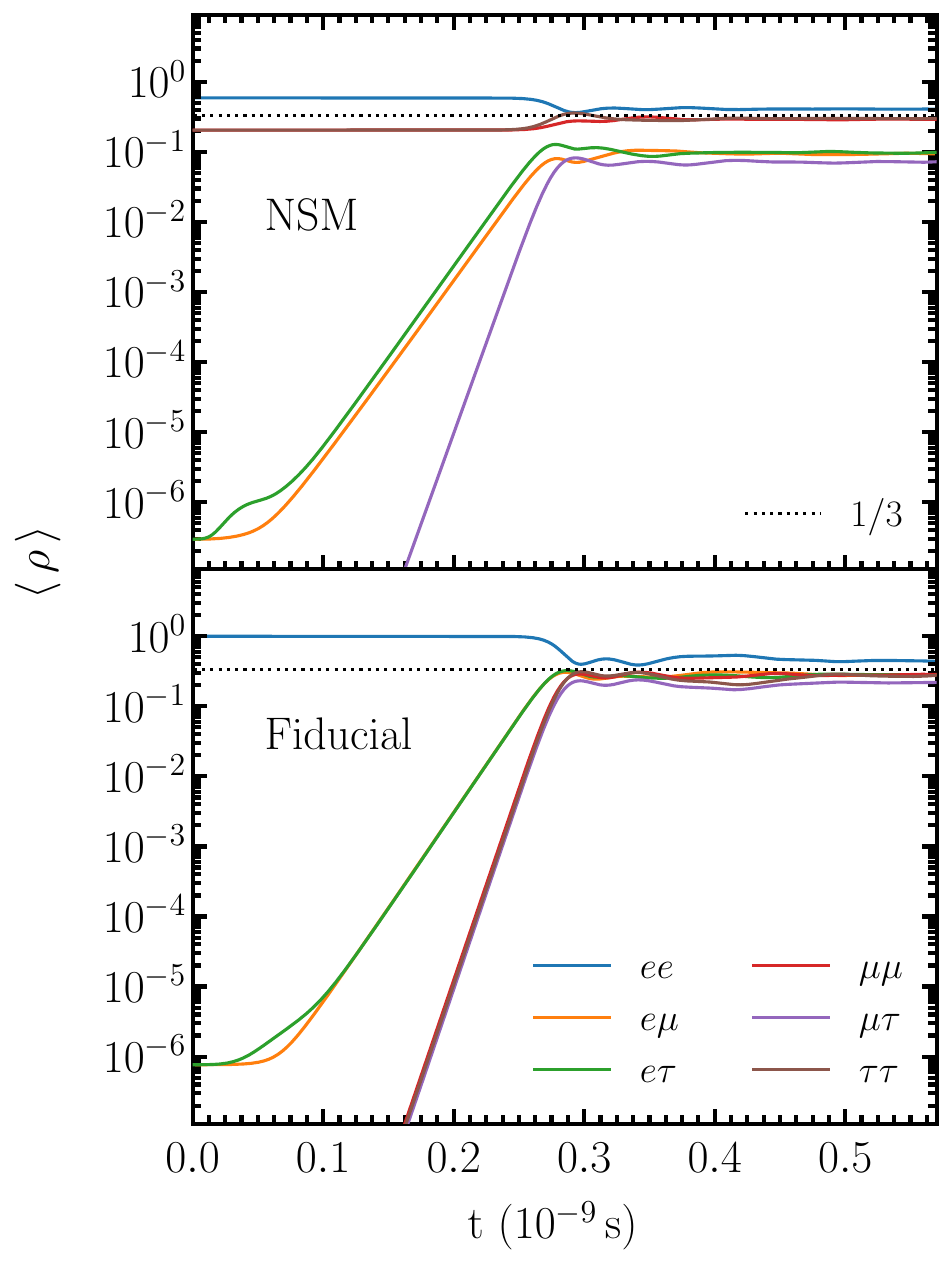}
            \caption{\label{fig: rho_domain_average} Density matrix averaged over the spatial domain of the NSM snapshot (upper panel) and Fiducial (lower panel) simulations. Individual components are visualized in distinct colors. In the off-diagonal components, flavor oscillation modes that start the simulation with small amplitudes grow, creating a combination of modes that stop the exponential trend due to unstable modes. The diagonal components reach a final state near complete flavor mixing (indicated by the black dotted line) and slightly fluctuate around it. The domain-averaged antineutrino density matrix follows the same trend.}
        \end{figure}
        
        To start, we need to first examine the impact of the fast flavor instability on the flavor distribution before delving into the additional effects introduced by chaos. This preliminary analysis will aid in interpreting the results presented in the subsequent sections. Figure \ref{fig: rho_domain_average} shows the neutrino density matrix averaged across the spatial domain, denoted as $\left<\,\rho\,\right>$. For flavor components $i,j \in\{\,e,\,\mu,\tau\}$, the domain-averaged density matrix is defined as a sum over the computational particles according to
        \begin{eqnarray}
            \left<\,\rho\,\right>_{ij}&=&\frac{\left<\,N\,\right>_{ij}}{\mathrm{Tr}\left<\,N\,\right>},
            \label{eq: domain_averaged_neutrino_density_matrix}
        \end{eqnarray}
        where
        \begin{eqnarray}
            \left<\,N\,\right>_{ij}&=&\frac{1}{N_{\mathrm{par}}}\sum_{k=1}^{N_{\mathrm{par}}} |N^{k} \rho^k_{ij}|.
        \end{eqnarray}
        Here, $N_\mathrm{par}$ represents the number of computational particles in the simulation, $N^k$ is the number of physical neutrinos represented by the computational particle $k$, and $\rho^k_{ij}$ denotes the neutrino density matrix of particle $k$.
        
        As previously reported in \cite{Particle-in-cell,three_dimensions,grohs2022neutrino}, the evolution of neutrino flavor exhibits three distinct phases: linear growth, saturation, and decoherence of flavor oscillation modes. During the linear growth phase, observed between $0.10$ and $0.25$ ns in both panels of Figure \ref{fig: rho_domain_average}, the off-diagonal components of the domain-averaged density matrix are predominantly influenced by highly unstable flavor oscillation modes that exhibit exponential growth as  $\left<\,\rho\,\right>_{ij}=A e^{-i\omega t}$ with $\mathrm{Im}(\omega) \neq 0$ and $A$ a small initial amplitude. The imaginary frequency of the dominant unstable modes can be read from the slopes of the off-diagonal curves during the linear growth phase. In the upper panel, the measured instability growth rates for the NSM simulation are $\mathrm{Im}(\omega_{e\mu})\approx 58.8 \,\mathrm{ns}^{-1}$ and $\mathrm{Im}(\omega_{e\tau})\approx 59.6 \,\mathrm{ns}^{-1}$. The bottom panel shows measured growth rates for the Fiducial simulation of $\mathrm{Im}(\omega_{e\mu})\approx 62.6 \,\mathrm{ns}^{-1}$ and $\mathrm{Im}(\omega_{e\tau})\approx 62.1 \,\mathrm{ns}^{-1}$. This is consistent with the previously reported values.
        
        After the linear growth phase, the flavor oscillation modes that begin the simulation with a negligible amplitude become noticeable, provoking saturation of the flavor oscillation modes.  This saturation phase occurs between $0.25$ and $0.30$ ns in both simulations. Significant flavor conversion does not take place until the conclusion of the linear growth phase and the initiation of the saturation phase. After the saturation phase, all the flavor oscillation modes start evolving in an incoherent mixture. This is the flavor decoherence phase. Coherent modes that grew in the linear growth phase are disrupted, and the phase of the oscillation modes is randomized, although long-lived modes can exist after the saturation phase \cite{duan_2022_flavor_isospin}. In the decoherence phase, neutrinos and antineutrinos start evolving in a complex non-linear relationship, archiving huge amounts of flavor conversion. After about $0.3\,\mathrm{ns}$, the domain-averaged neutrino density matrix in the NSM simulation reaches an incomplete flavor mixing state as an equilibrium point and fluctuates slightly around that state. The domain-averaged antineutrino density matrix follows the same pattern. The dynamics of the Fiducial simulation are remarkably similar to that of the NSM simulation, except that the equilibrium distribution of flavors is very close to an even mixture.
        
    \subsection{Chaos}
    \label{sec:results:chaos}
    
        To quantify the evolution of the small perturbation $\vec{\delta}_t$ over time, we introduce a mathematical vector henceforth referred to as the flavor vector $\vec{r}$, with components given by
        \begin{eqnarray}
            r_{16k+l}&=&N^kP^k_l.
            \label{eq: flavor_vector}
        \end{eqnarray}
        Here, $k\in \{0,N_\mathrm{par})$ denotes the particle index, $l \in [0, 7]$ represents the Gell-Mann vector component of a single particle's neutrino flavor vector, and $l \in [8,15]$ designates the Gell-Mann component for a particle's antineutrino flavor vector. $P^k_{l}=\frac{1}{2}\mathrm{Tr}(\rho^k G_l)$ is the neutrino flavor polarization vector for the computational particle $k$, where $G_l$ refers to the Gell-Mann matrices (with the eight Gell-Mann matrices being repeated for $l \in [8,15]$). Given that $l \in [0,15]$, the flavor vector has $16N_{\mathrm{par}}$ components. The flavor vector completely determines the flavor content of our simulations and its evolution can be thought of as tracing a path through a $16N_\mathrm{par}$-dimensional state space. Given that the length of the flavor polarization vector of each particle is a constant of motion, the trajectories of the flavor vector in the state space is bounded to the surface of a $16N_\mathrm{par}$-dimensional sphere of radius the length of the flavor vector (see Appendix \ref{appendixb}).
        
        The evolution over time of small perturbations can be obtained by simulating the paths in the state space of two flavor vectors with initial conditions $\vec{r}_{\mathrm{bas}}(t_0)=\vec{r}_{t_0}$ and $\vec{r}_{\mathrm{per}}(t_0)=\vec{r}_{t_0}+\vec{\delta}_{t_0}$. Here, $\vec{r}_{t_0}$ is the initial flavor vector of the NSM and Fiducial simulations described in Sections \ref{Initial conditions NSM snapshot simulation} and \ref{Initial conditions: Fiducial simulation}, respectively. The perturbation $\vec{\delta}_{t}$ can be obtained by computing 
        \begin{eqnarray}
            \vec{\delta}_{t}=\vec{r}_{\mathrm{per}}-\vec{r}_{\mathrm{bas}}.
        \end{eqnarray}
        The first simulation, described by $\vec{r}_{\mathrm{bas}}$ serves as the baseline simulation for comparison. The other, described by $\vec{r}_{\mathrm{per}}$, is randomly perturbed from the baseline simulation. Specifically, we randomly choose the initial magnitude of the perturbation so that
        \begin{eqnarray}
            \frac{|\vec{\delta}_{t_0}|}{\left|\vec{r}_{t_0}\right|}\sim 10^{-10}.
        \end{eqnarray}
        This choice of perturbation is sufficiently large to avoid numerical errors (see Appendix \ref{Convergence_test}). The mathematical maximum magnitude of the perturbation at any point in time is $2|\vec{r}_{t}|$ (i.e., when $\vec{r}_\mathrm{per}$ is oriented opposite to $\vec{r}_\mathrm{bas}$). It is important to note that the "chaos" perturbation is independent of the perturbation used to seed the instability (see Section~\ref{sec:methods}), which is present in both the baseline and perturbed simulations. In the following discussion, the perturbations referred to are the "chaos" perturbations and not the initial instability seed.
        
        \begin{figure*}[!htbp]
        \includegraphics[width=0.95\textwidth]{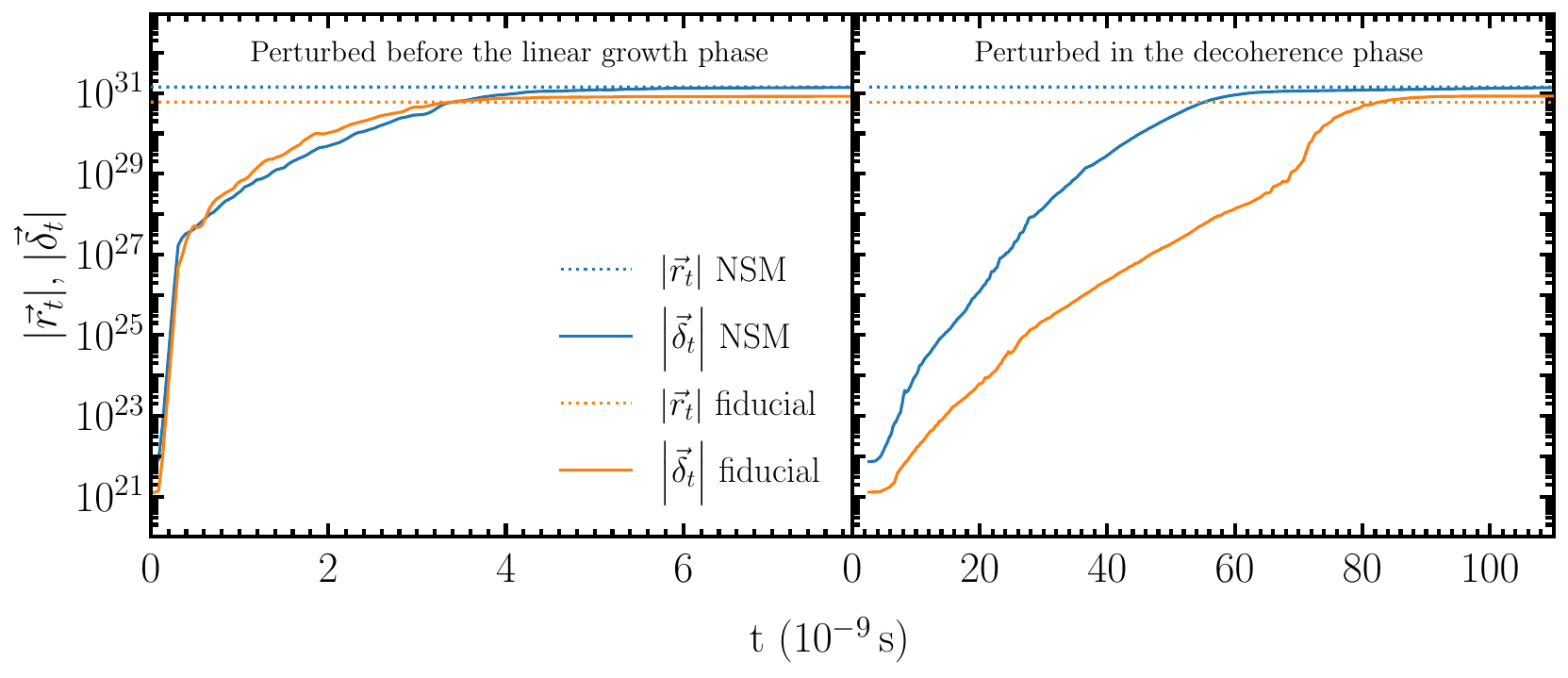}
        \caption{\label{fig: delta_state_space_exponential_divergence}
        Time evolution of small perturbations $|\vec{\delta}_{t}|$ in the NSM (blue) and Fiducial (orange) simulations. In the left panel, the perturbation is applied at $t_0=0\, \mathrm{ns}$ (before the linear growth phase). Between $0.10$ and $0.25$ ns, the perturbation growth is exponentially driven mainly by the fast flavor instability. In the right panel, to avoid the presence of unstable fast flavor modes, the perturbation is applied in the decoherence phase at $t_0=2.65\, \mathrm{ns}$. The magnitude of the perturbation grows, following an approximate exponential trend. This demonstrates that paths of similar flavor vectors in the state space diverge exponentially, illustrating the chaotic evolution of neutrino flavor. As the simulation concludes, the magnitude of the perturbations approaches a constant value larger than the magnitude of the flavor vector $|\vec{r}_{t}|$ (dotted lines).
        }
        \end{figure*}
        
        If the neutrino flavor evolution is chaotic, the time evolution of small perturbations follows an approximately exponential trend (see Equation \ref{eq: exponential_trend_of_perturbations}) with $\lambda>0$. To test whether perturbations exhibit this exponential behavior, we performed two simulations in which we apply perturbations at $t_0=0$ and $t_0=2.65$ ns in both the NSM and Fiducial simulations. The first case represents a perturbation applied before the linear growth phase, and the second a perturbation applied in the decoherence phase.
        
        The blue curve on the left panel of Figure \ref{fig: delta_state_space_exponential_divergence} shows the evolution of a perturbation applied at $t_0=0$ in the NSM simulation. During the linear growth phase, between $0.10$ and $0.25$ ns, the perturbation follows an exponential trend characterized by $\lambda\approx 59$ ns$^{-1}$. This trend is mainly driven by fast flavor unstable modes, and does not represent a signature of chaos. Additionally, as previously exposed, there is no significant flavor conversion at this time. Towards the end of the linear growth phase, the perturbation's exponential trend abruptly transitions to a slower rate ($\lambda \approx 2.6$ ns$^{-1}$). At this point, as observed in Figure \ref{fig: rho_domain_average}, unstable modes have vanished and the domain-averaged density matrix evolves into a complex, incoherent mixture of flavor oscillation modes that has attained an equilibrium distribution, closely resembling flavor equipartition. This suggests that the exponential growth of the perturbation in decoherence phase is not driven by fast flavor unstable modes and is a signature of chaotic flavor evolution in the flavor vector. The Fiducial simulation (orange) shows a very similar trend for the same reasons. The growth rates before and after saturation in this case are approximately $58$ ns$^{-1}$ and $2.6$ ns$^{-1}$, respectively.
        
        To isolate the chaotic behavior from the fast flavor instability, we perform another simulation where we apply the perturbation at $t_0=2.65\,\mathrm{ns}$ during the decoherence phase (right panel of Figure \ref{fig: delta_state_space_exponential_divergence}). In the NSM simulation (blue), the perturbation amplitude exhibits an approximately exponential growth with $\lambda=0.44\,\mathrm{ns}^{-1}$, and in the Fiducial simulation (orange), the growth rate is $\lambda=0.32\,\mathrm{ns}^{-1}$. This observation highlights that the paths of similar flavor vectors in the state space diverge exponentially, indicating chaotic behavior. The Lyapunov exponent is one order of magnitude smaller than the growth rate of the dominant flavor unstable modes present in the linear growth phase.
        
        \begin{figure*}[!htbp]
            \includegraphics[width=0.8\textwidth]{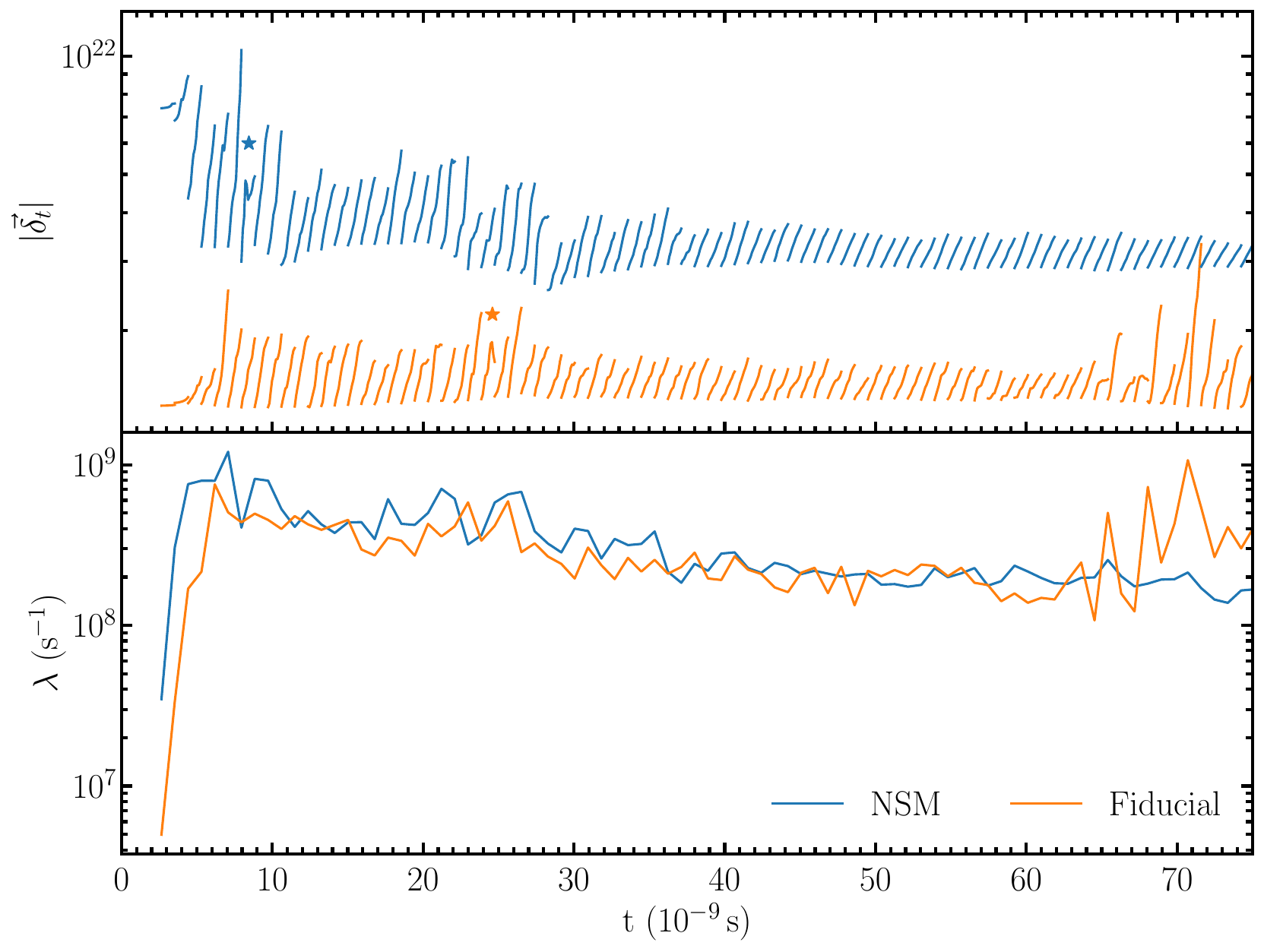}
            \caption{\label{fig: r_periodic_renormalization} The upper panel shows the time evolution of perturbations periodically normalized to keep the perturbation small. The blue curve corresponds to the NSM simulation, while the orange curve represents the Fiducial simulation. The lower panel shows the Lyapunov exponent calculated over each interval between normalizations. The Lyapunov exponent of a perturbation in the state space varies based on the direction of the perturbation. The presence of exponential divergence ($\lambda>0$) and convergence ($\lambda<0$ e.g. below the blue and orange stars) is an expected result of the Liouville theorem of conservation of the state space volume of canonical variables.    
            }
        \end{figure*}
        
        It is also curious that the growth rates for perturbations applied before the linear growth phase are so different from those applied in the decoherence phase. The Lyapunov exponent depends on the position in the state space of both baseline and perturbed flavor vectors (i.e., the direction of the $16N_{\mathrm{par}}$-dimensional perturbation vector $\vec{\delta}_{t}$), so as the system evolves $\lambda$ can change, and a single simulation is not able to fully explore such a large parameter space. To illustrate this, we run a third set of simulations in which we periodically normalize the perturbation maintaining its direction every $0.9$ ns. The upper panel of Figure \ref{fig: r_periodic_renormalization} shows the time evolution of the perturbation after each normalization, while the lower panel provides an approximation of the Lyapunov exponents calculated over each interval between normalizations. Given the significant differences in the initial conditions, it is remarkable how similar chaos manifests in the NSM and Fiducial simulations. The Lyapunov exponents emerging from the pre-saturation perturbations are larger because the fast flavor instability causes the perturbation to grow in a very specific direction that is closer to directions that exhibit the strongest chaos. Randomly perturbing after the instability saturates simply sets the stage with a random perturbation in a less chaotic direction. In bipolar simulations involving only two beams it is possible to obtain the full spectrum of Lyapunov exponents \cite{Hansen_2014}, but that is infeasible with these large-scale simulations. However, we do find that random perturbations produce consistent Lyapunov exponents across multiple runs, leading us to believe that the numbers reported for the post-saturation perturbations represent a lower bound for the chaoticity of the system.
        
        Although the exponential trend of the perturbations is predominantly driven by positive Lyapunov exponents ($\lambda>0$), the upper panel reveals the existence of directions with negative Lyapunov exponents ($\lambda<0$). One such direction is marked by each a blue and orange star. Stable directions with negative Lyapunov exponents are swiftly suppressed by divergent directions. In other words, perturbations that initiate evolution in a stable direction will eventually transition to a direction characterized by a positive Lyapunov exponent. This is another illustration of the dependence of the Lyapunov exponent on the state itself and the direction of the perturbation.
        
        Negative Lyapunov exponents arise from the conservation of the state space volume of the canonical variables, as stated in the Liouville theorem. The QKE can be transformed into a conservative classical Hamiltonian system with the canonical coordinates and momenta given by any function of the coordinates used to describe the flavor polarization vector (see Section II A of \cite{Hansen_2014}). This implies that for each direction in the state space with exponential divergence, there will be a direction of exponential convergence at the same rate, ensuring the conservation of the state space volume of the canonical variables, even if its shape is deformed. The exponential divergence of small perturbations also has consequences for the integrability of the QKE. In Hamiltonian systems, for each conserved quantity, there exists one direction in the state space with a zero Lyapunov exponent. Since integrable systems have conserved quantities as degrees of freedom, the Lyapunov exponent is zero for every direction in the state space. In other words, the shape of the state space volume of the canonical variables remains unchanged. Even though our state space does not correspond to the state space of the canonical variables, our finding that close flavor vector paths in state space diverge exponentially suggests that the QKE is a non-integrable system.
        
        Towards the end of the simulation, the magnitude of the perturbation stabilizes, reaching a constant value ($t \gtrsim 4 \, \mathrm{ns}$ in the left panel and $t \gtrsim 60 \, \mathrm{ns}$ in the right panel). In both cases, the perturbation's magnitude asymptotically settles to a value less than the maximum possible value of $2|\vec{r}_{t}|$ but greater than $|\vec{r}_{t}|$. If we interpret $\vec{\delta}_{t}$ as the uncertainty in the flavor vector $\vec{r}_{t}\pm\vec{\delta}_{t}$, the relative error in $\vec{r}_{t}$ can be up to $106\%$ for the NSM simulation, and 142\% for the Fiducial simulation. This implies that even small uncertainties are exponentially amplified making the flavor vector unpredictable on a time scale of the inverse of the Lyapunov exponent. The Fiducial simulation maximum perturbation amplitude is considerably larger than in the NSM simulation because the Fiducial simulation experiences a higher amount of flavor conversion in all directions. All particles in the Fiducial simulation change flavor from a pure electron state to a flavor-mixing equilibrium.

    \subsection{Individual Particles}
    \label{Individual Particles}
        
        \begin{figure}[!htbp]
            \includegraphics[width=0.95\linewidth]{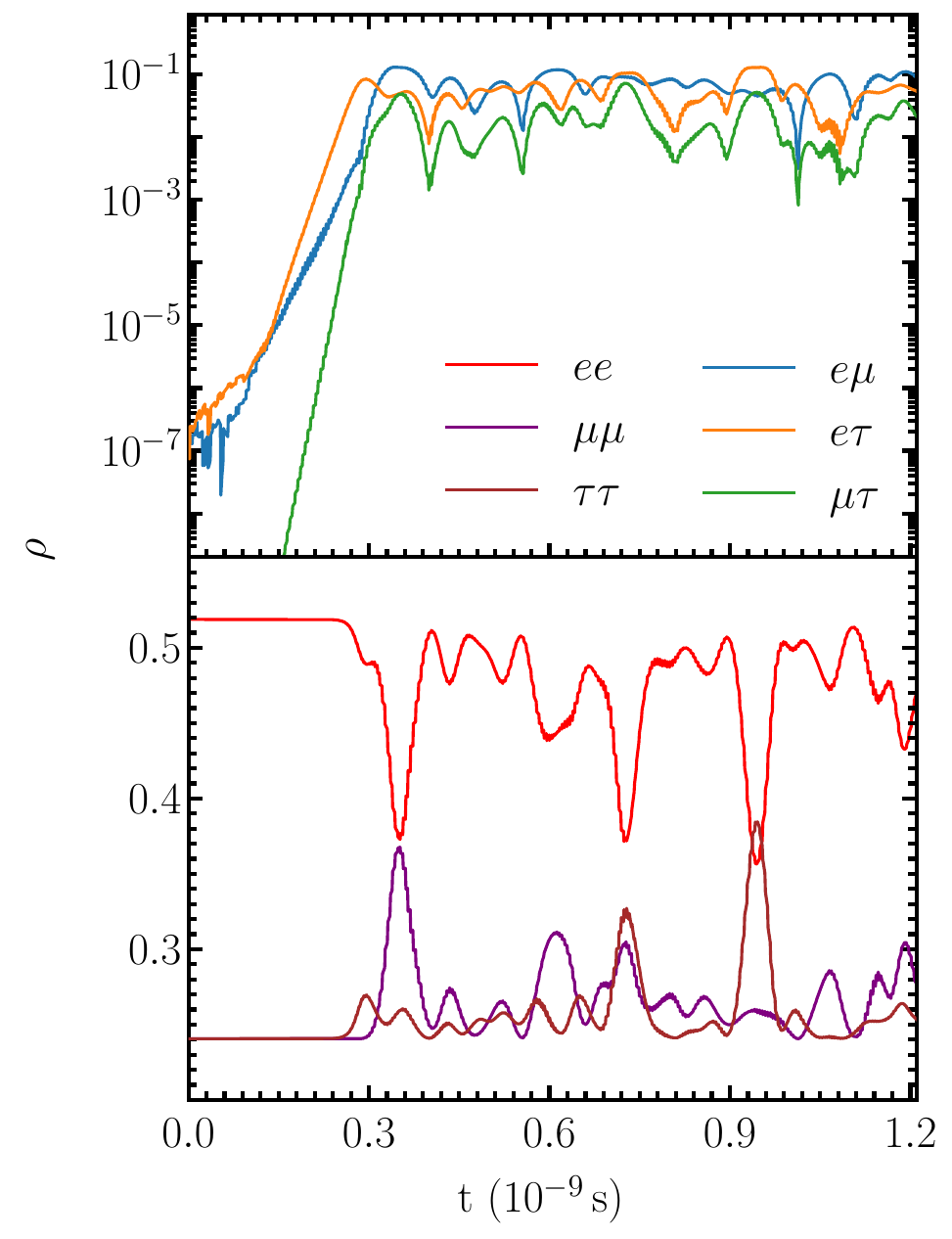}
            \caption{\label{fig: rho_single} The upper panel shows the magnitude of the non-diagonal components of the neutrino density matrix for a single computational particle in EMU, obtained from the NSM snapshot simulation. The lower panel shows the diagonal components of the neutrinos density matrix for the same computational particle. The flavor transformation becomes evident at the end of the linear growth phase and evolves with large amplitude, in contrast to the domain-averaged quantities in Figure \ref{fig: rho_domain_average}.}
        \end{figure}
        
        While the domain-averaged density matrices tend to a flavor mixing equilibrium, this stationary state comprises numerous particles whose flavor states undergo random and incoherent fluctuations at small scales. The fine details of these fluctuations, observable at the level of single computational particles, remain highly chaotic.
        
        This can be seen in figure \ref{fig: rho_single}, which shows the neutrino flavor transformation of a single computational particle in the EMU code for the NSM simulation. The upper panel shows the magnitude of the off-diagonal components of the neutrino density matrix, while the lower panel displays the diagonal components. Similar to Figure \ref{fig: rho_domain_average}, the same linear growth, saturation, and decoherence phases are discernable. Flavor transformations become evident toward the end of the linear growth phase, around $0.25$ ns. The amount of flavor conversion that a computational particle experiences is related to the direction of its momentum. The flavor transformation follows a complex non-linear trend with no discernible pattern. 
        
        \begin{figure}[!htbp]
            \includegraphics[width=0.95\linewidth]{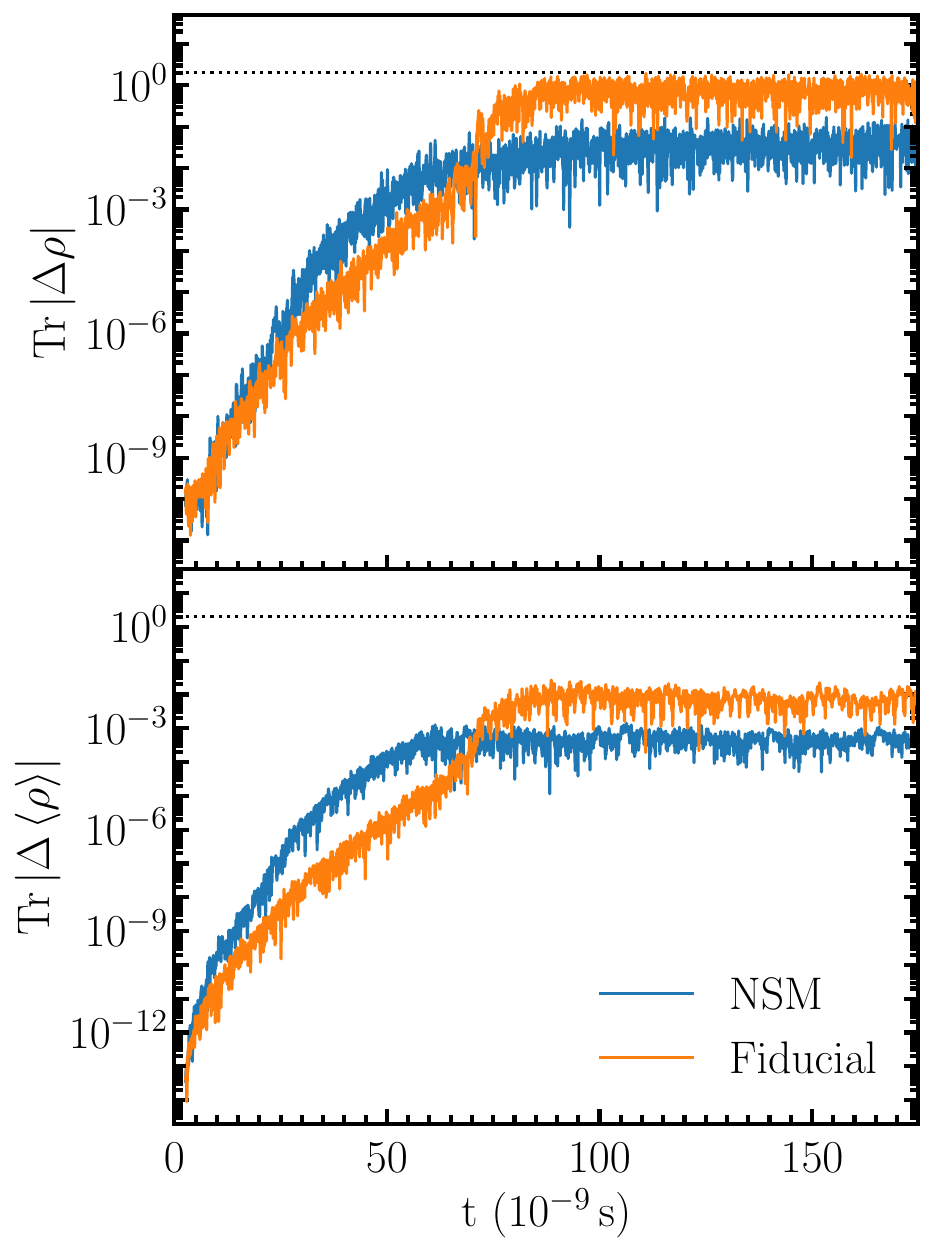}
            \caption{\label{fig: difference_rho} The upper panel shows the flavor-traced difference between the neutrino density matrices (Equation~\ref{eq: trace_rho}) of single computational particles extracted from the baseline and post-saturation perturbed simulations. The lower panel shows the same quantity, but using the domain-averaged density matrix $\langle \rho \rangle$. The blue and orange curves denote the NSM and Fiducial simulations, respectively. The black dotted lines show the maximum possible value. Both panels exhibit a comparable exponential trend, akin to the perturbations observed in the flavor vectors (see the right panel of Figure \ref{fig: delta_state_space_exponential_divergence}), but late-time uncertainties in the domain-averaged quantities are much smaller than in individual particles.}
        \end{figure}
        
        The Lyapunov exponent extracted from individual particles is similar to those obtained from the full state vector in Section~\ref{sec:results:chaos}. The blue (NSM) and orange (Fiducial) curves in the upper panel of Figure \ref{fig: difference_rho} show the difference between the flavor state of a single particle extracted from the baseline simulations and the simulations perturbed at $2.65\,\mathrm{ns}$. To avoid focusing on a single flavor, we sum over flavors and plot
        \begin{eqnarray}
            \mathrm{Tr}| \Delta \rho|=| \Delta \rho|_{\mathrm{ee}}+| \Delta \rho|_{\mu\mu}+| \Delta \rho|_{\tau\tau}.  
            \label{eq: trace_rho}
        \end{eqnarray}
        This can be interpreted as the combined uncertainty in the average electron, muon, and tau flavors generated by the perturbation in the flavor vector, whose maximum value is two. This difference follows a similar exponential trend ($\lambda\approx 0.41$ ns$^{-1}$ for NSM and $0.32$ ns$^{-1}$ for Fiducial) as the full flavor vector (right panel of Figure \ref{fig: delta_state_space_exponential_divergence}). Toward the end of the simulation, this difference ceases exponential growth and instead shows irregular and large-amplitude fluctuations. The maximum combined uncertainty reaches values of up to $0.17$ for the NSM particle and $1.92$ for the Fiducial particle. By tracking particles propagating in other directions (not shown), we note that computational particles experiencing high-amplitude flavor oscillation modes reach values of $\mathrm{Tr}|\Delta\rho|$ close to two (similar to the Fiducial particle), while particles undergoing low-amplitude flavor oscillation modes exhibit small values of $\mathrm{Tr}|\Delta\rho|$ (like the NSM particle).
        
    \subsection{Domain-Averaged Quantities}
    \label{Domain-Averaged Quantities}
        
        The flavor vector in Equation \eqref{eq: flavor_vector} fully characterizes a many-body flavor quantum state (which by construction never builds multi-particle entanglement). However, in astrophysical applications, such as CCSNe and NSMs, this level of detail is unnecessary and computationally infeasible. A more relevant quantity is the domain-averaged density matrix over a spatial domain greater than the flavor oscillation length scale. The quantification of the impact of chaos on these quantities is important for a reliable implementation of the neutrino flavor transformation in CCSN and NSM simulations, where even a small numerical error could propagate exponentially.
        
        To investigate how small perturbations propagate on macroscopic scales, we compute the domain-averaged density matrices defined in Equation \eqref{eq: domain_averaged_neutrino_density_matrix} for the baseline simulations and the simulations perturbed at $2.65\,\mathrm{ns}$ used in the right panel of Figure \ref{fig: delta_state_space_exponential_divergence}. The lower panel of Figure \ref{fig: difference_rho} shows the time evolution of the trace of the difference between the baseline and perturbed neutrino density matrices averaged in the domain, defined as in Equation \eqref{eq: trace_rho} but using the domain-averaged density matrix $\langle \rho \rangle$. In the NSM simulation the maximum value it reaches represents $0.07\%$ of the maximum possible value of two, while in the fiducial simulation it represents $1.3\%$ of the maximum. This is considerably smaller than the relative error of the flavor vector due to the same perturbation, implying that small perturbations do not escalate to large magnitudes in domain-averaged quantities. The diagonal components of the domain-averaged density matrix have a high degree of predictability compared to the flavor vector even amidst chaotic flavor evolution. Due to this distinctive feature, (i.e., small initial perturbations do not reach large magnitudes) it is reasonable to use the domain-averaged density matrix as a key variable in simulations of neutrino flavor transformation in CCSNe and NSMs. This is also encouraging for thermodynamic theories of flavor transformation (e.g., \cite{johns2023thermodynamics}).

\section{Conclusions}
\label{sec:conclusions}

    To inform simulations of neutrino flavor transformation in CCSNe and NSMs, we delve into the chaotic nature of neutrino flavor transformations in two distinct one-dimensional dense neutrino gases. In the first simulation (Section~\ref{Initial conditions NSM snapshot simulation}), we extract the neutrino distribution from a region above the accretion disk of a multidimensional NSM simulation, and in the second simulation (Section~\ref{Initial conditions: Fiducial simulation}), we examine a well-understood toy model. Both distributions undergo significant fast flavor instabilities (see Section \ref{Overall Dynamics}).
    
    Using Lyapunov exponents, we study the stability of paths in the state space of similar flavor vectors (Equation \ref{eq: flavor_vector}) with similar initial conditions. These paths diverge exponentially (see Section~\ref{sec:results:chaos}) showing that the neutrino flavor transformation is chaotic.  Interpreting the perturbation as uncertainty results in a final relative error of over $100$ percent in the position of the flavor vector in the state space that grows with a Lyapunov exponent with a lower limit of $0.44\,\mathrm{ns}^{-1}$ (NSM) or $0.32\,\mathrm{ns}^{-1}$ (Fiducial), as shown in Figure \ref{fig: delta_state_space_exponential_divergence}. Since the flavor vector completely characterizes the flavor content in the simulations, we conclude that the long-term evolution of neutrino and antineutrino flavors at this level of detail is unpredictable.
    
    We analyze how the Lyapunov exponents depend on the direction in the state space of the perturbation (see Figure \ref{fig: r_periodic_renormalization}). In both simulations, we find that the time evolution of perturbations is mainly driven by directions of positive Lyapunov exponents. A perturbation applied in a stable direction ($\lambda<0$) will eventually move to an unstable direction in state space ($\lambda>0$).
    
    We identify stable directions in state space in which the perturbations briefly converge exponentially below the stars in the upper panel of Figure \ref{fig: r_periodic_renormalization}. Since the QKE can be transformed into a classical Hamiltonian system (as in Equation 2 of \cite{Hansen_2014}) several Hamiltonian mechanic theorems can be applied. The Liouville theorem claims that the state space volume of the canonical coordinates is a constant of motion, i.e., the spectrum of the Lyapunov exponents is symmetric
    \begin{eqnarray}
        (\lambda_1,\,\lambda_2,\,\lambda_3,\,...\,,-\lambda_3,\,-\lambda_2,\,-\lambda_1)\,\,,
    \end{eqnarray}
    meaning that for each direction in the state space with exponential divergence, there exists another direction of exponential convergence at the same rate, so the volume of the state space is conserved even though its shape is deformed. One Lyapunov exponent is zero for each conserved quantity, and all of them are zero for a stable or integrable system where there exist conserved quantities as degrees of freedom. The presence of a direction of exponential divergence and convergence in our simulations suggests that the QKE forms a non-integrable system.
    
    The unpredictable nature of chaos manifests differently on the macroscopic and microscopic scales of neutrino flavor transformation. The macroscopic scale is represented by the domain-averaged density matrix, while the microscopic scale is the quantum state of single computational particles that represents the smallest spatial resolution of flavor in the simulations. On single computational particles, small perturbations grow exponentially (see Section~\ref{Individual Particles}), reaching magnitudes of order unity for particles moving in angular directions with a large amount of flavor conversion. This implies a relative error of $100\%$ on the quantum states of neutrinos and antineutrinos, meaning that slightly different initial flavor vectors reproduce a different flavor evolution in single computational particles. Individual computational particles destroy information at a rate of $\lambda \log_{2}{e} \approx 0.6$ bits/ns, implying that initial variables specified with 64-bit floating-point precision (of which only 53 bits are used to store significant digits) cannot be accurately predicted after approximately $100$ ns.
    
    In the domain-averaged density matrix, small perturbations in the flavor vector grow exponentially reaching small magnitudes compared to the maximum possible value. In both simulations, this produces a combined maximum uncertainty in the diagonal components of the domain-averaged density matrix of less than $1.3\%$ (see Section~\ref{Domain-Averaged Quantities}). This suggests that CCSN and NSM simulations could safely rely on domain-averaged quantities even if microscopic details are unpredictable.

\section{Acknowledgements}
    We thank Zewei Xiong and Meng-Ru Wu for useful discussions. EU was supported by the Dr. Elizabeth M. Bains and Dr. James A. Bains Graduate Fellowship. SR was supported by a National Science Foundation Astronomy and Astrophysics Postdoctoral Fellowship under award AST-2001760. This research used resources of the National Energy Research Scientific Computing Center, a DOE Office of Science User Facility supported by the Office of Science of the U.S. Department of Energy under Contract No. DE-AC02-05CH11231 using NERSC award NP-ERCAP0021631. 
    
\appendix

\section{Convergence tests}
\label{Convergence_test}
    
    We are concerned with the numerical convergence of the Lyapunov exponent. These convergence tests are conducted for the Fiducial simulation but can be generalized to the NSM simulation since the numerical scheme in the EMU code remains the same and both simulation exhibit similar dynamics and numerical requirements.
    
    Particle-in-cell neutrino simulations in EMU rely on three imposed parameters: the domain size, the number of cells, and the number of particles. These parameters must be carefully chosen to ensure convergence in the physical quantities under study. Another crucial aspect in computing the Lyapunov exponent is the magnitude of the perturbation. A small perturbation can lead to subtractive cancellation errors, while large magnitudes result in saturation before a useful trend can be established. A balance between these conditions needs to be achieved.
    
    In the following sections, we examine the impact of the domain size, the number of cells, the number of particles, and the magnitude of the perturbation on the Lyapunov exponent. This establishes the parameters for which the simulation accurately reproduces the Lyapunov exponent, minimizing numerical errors. To approximate the Lyapunov exponent in Equation \eqref{eq: exponential_trend_of_perturbations}, we employed the least squares linear fit method for $y=mx+b$ using the \texttt{numpy.polyfit()} function. Here, $y=\ln |\vec{\delta}_{t}|$, $m=\lambda$, $x=t$ and $b=\ln |\vec{\delta}_{t_0}|$. The \texttt{numpy.polyfit()} function returns values of $m$ and $b$ that minimize the sum of the squares of the residuals.

    \subsection{Number of particles}

        To determine whether the Lyapunov exponent is affected by the number of particles, we conducted five simulations wherein neutrinos are emitted in a roughly isotropic distribution from the center of each cell. We considered $92$, $378$, $6,022$, $24,088$, and $54,202$ particles per cell for each case. The simulation domain is $1\times1\times64$ cm, divided into $1024$ cells in the $\hat{z}$ direction.
        
        Starting at $t_0=2.65$ ns, we applied the perturbation in a random orientation with magnitude $| \vec{\delta}_{t_0} |/| \vec{r}_{t_0} |\sim 10^{-10}$. We evolved the perturbation until the simulation reached $20$ ns or until $| \vec{\delta}_{t} |/| \vec{r}_{t} |=10^{-6}$.
        
\begin{figure}[!htbp]
            \centering
            \includegraphics[width=1\linewidth]{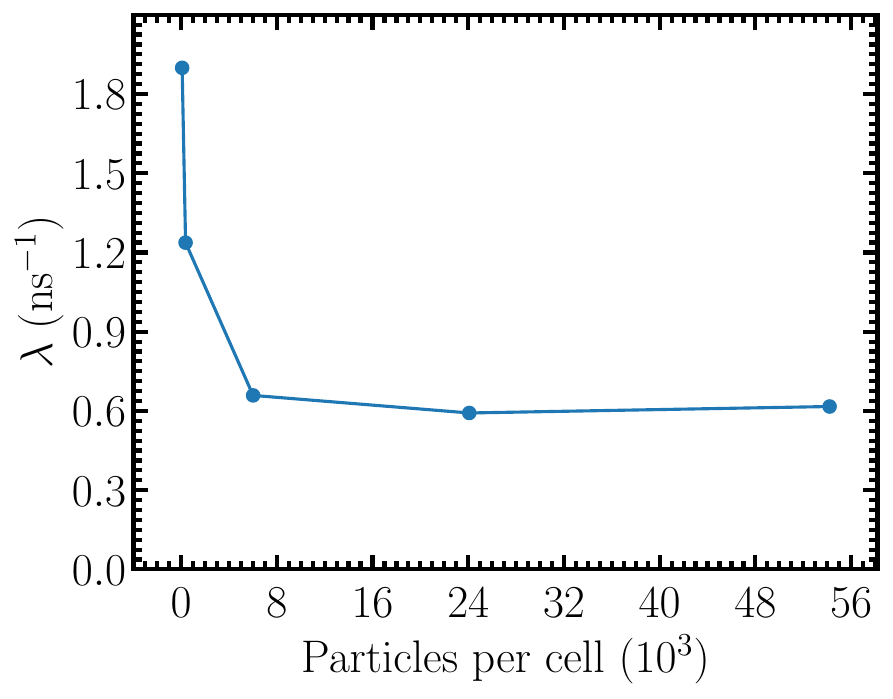}
            \caption{Test for convergence of the Lyapunov exponent with the number of particles in the EMU simulations. The vertical axis shows the Lyapunov exponent for five Fiducial simulations. The simulation domain is $1 \times 1\times 64$ cm divided into $1024$ cells. Convergence is attained for a number of directions greater than $24,088$.}
            \label{fig: convergence_number_of_particles}
        \end{figure}

        Figure \ref{fig: convergence_number_of_particles} shows the Lyapunov exponents for the five simulations in this test. As the number of particles (and hence the number of directions) increases, the Lyapunov exponent asymptotes to $0.6$ ns$^{-1}$. Around $24,088$ directions, convergence is achieved. This represents a high-precision simulation compared to the $92$ particles needed to achieve convergence in the long-term domain-averaged density matrix \cite{Particle-in-cell}.
        
    \subsection{Perturbation magnitude}
    
        To evaluate how the initial magnitude of the perturbation affects the Lyapunov exponent, we perform three sets of five simulations with initial perturbation magnitudes $| \vec{\delta}_{t_0} |/| \vec{r}_{t_0} |$ on the order of $10^{-14}$, $10^{-12}$ and $10^{-10}$. The simulation domain is $1\times 1\times 64$ cm divided by $1024$ cells. Each cell is initialized with $24,088$ particles at the center of the cell and the perturbation is applied at $t=2.65$ ns. We terminate the simulation at $20$ ns or until $|\vec{\delta}_{t} |/| \vec{r}_{t} |=10^{-6}$.
        
\begin{figure}[!htbp]
            \centering
            \includegraphics[width=1\linewidth]{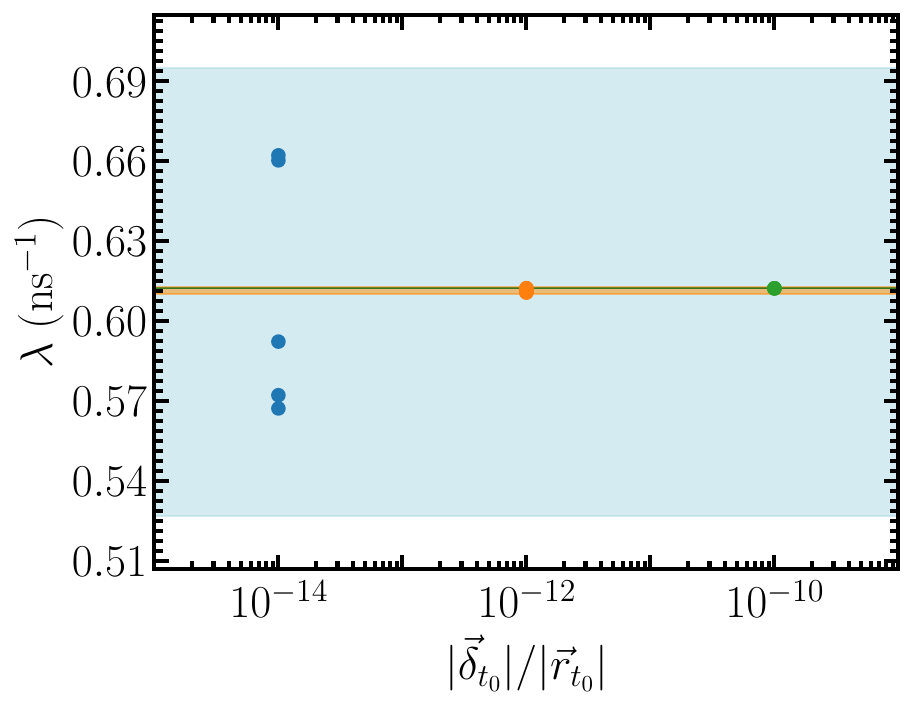}
            \caption{Test for convergence of the Lyapunov exponent with the magnitude of the perturbation. The vertical axis shows the Lyapunov exponent (blue, orange and green) for three sets of five simulations with distinct initial perturbation magnitudes ($10^{-14}$,$10^{-12}$, and $10^{-10}$). The simulation domain is $1\times1\times64$ cm, divided into $1024$ cells with $24,088$ particles per cell. The shaded areas represent a $2\sigma$ spread in the Lyapunov exponents of the same color. Convergence is attained for $| \vec{\delta}_{t_0} |/| \vec{r}_{t_0} |\sim 10^{-10}$.}
            \label{fig: convergence_magnitude_of_perturbation}
        \end{figure}

        Figure \ref{fig: convergence_magnitude_of_perturbation} show the Lyapunov exponent (blue, orange and green) for three sets of five simulations with distinct initial perturbation magnitudes ($10^{-14}$,$10^{-12}$, and $10^{-10}$). The blue, orange, and green shaded areas represent a spread of $2\sigma$ in the Lyapunov exponents of the same color. Larger perturbation magnitudes result in less dispersion in the Lyapunov exponent without significantly changing the mean ($0.61$ ns$^{-1}$), confirming their independence from the perturbation magnitude. At a perturbation magnitude $| \vec{\delta}_{t_0} |/| \vec{r}_{t_0} |\sim 10^{-10}$, the spread of in the Lyapunov exponent is negligible ($\sigma \approx 6.2\times 10^{-6}$ ns$^{-1}$). This can be considered a convergence point.
        
    \subsection{Domain size}
    
        \begin{figure}[!htbp]
            \centering
            \includegraphics[width=1\linewidth]{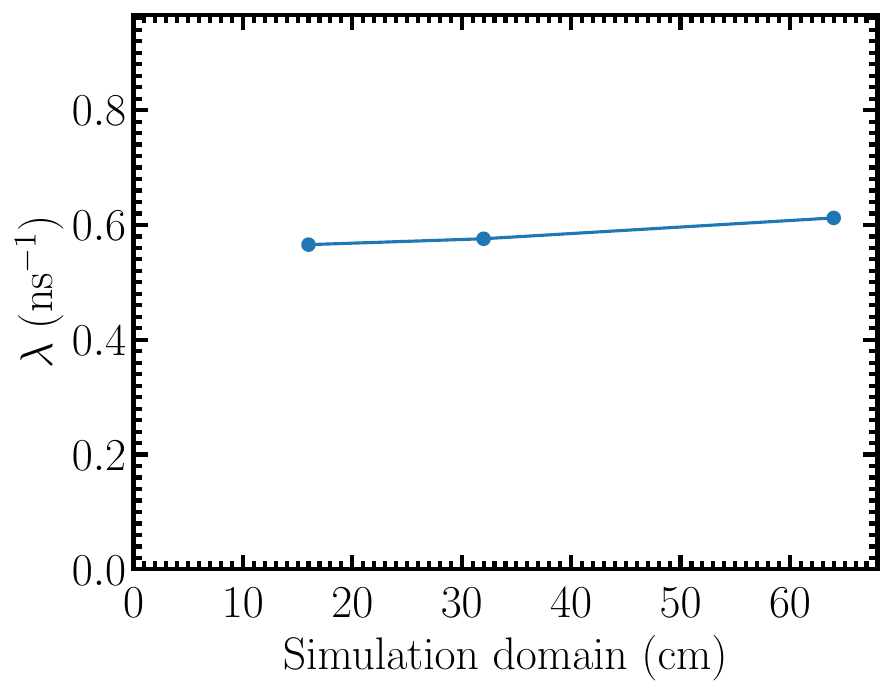}
            \caption{Test for convergence of the Lyapunov exponents with the simulation domain size. The vertical axis shows the Lyapunov exponents for three simulations with domain sizes of $16$, $32$, and $64$ cm in the $\hat{z}$ direction and $1$ cm in the $\hat{x}$ and $\hat{y}$ directions. The $\hat{z}$ direction has $16$ cells per centimeter, while the $\hat{x}$ and $\hat{y}$ directions each have one cell per centimeter. Each cell contains $24,088$ particles. The Lyapunov exponents remain relatively stable as the simulation size increases.}
            \label{fig: convergence_domain_size}
        \end{figure}
        
        To investigate the impact of domain size on the Lyapunov exponent, we conducted three simulations with dimensions of $16$, $32$, and $64$ cm in the $\hat{z}$ direction and $1$ cm in the $\hat{x}$ and $\hat{y}$ directions. There are $16$ cells per centimeter in the $\hat{z}$ direction and one in the $\hat{x}$ and $\hat{y}$ directions. In each cell, $24,088$ particles are emitted in an approximately isotropic distribution from the center. A perturbation with $| \vec{\delta}_{t_0} |/| \vec{r}_{t_0} |\sim 10^{-10}$ is applied at $t=2.65$ ns long after the saturation of flavor instability. We allowed the perturbation to evolve until the simulation reached $20$ ns or until $| \vec{\delta}_{t} |/| \vec{r}_{t} |\sim 10^{-6}$.
        
        Figure \ref{fig: convergence_domain_size} shows the dependence of Lyapunov exponents on the size of the simulation domain. The Lyapunov exponents do not exhibit significant variation and remain approximately constant as the simulation domain increases. We consider convergence achieved for a simulation domain of $1\times 1 \times64$ cm.
        
    \subsection{Cell size}
    
        \begin{figure}[!htbp]
            \centering
            \includegraphics[width=1\linewidth]{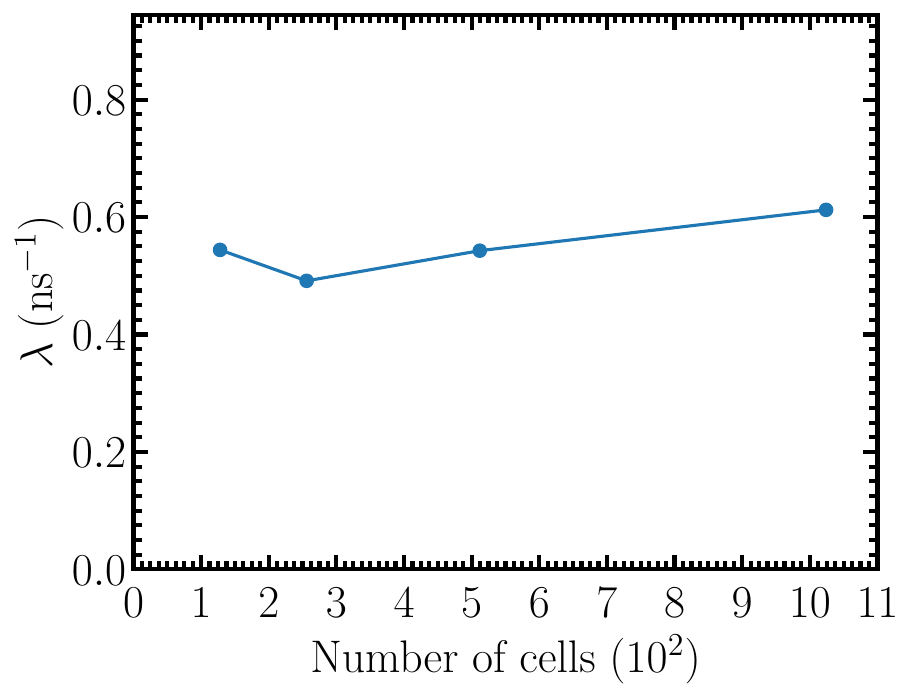}
            \caption{Test for convergence of the Lyapunov exponents with the cell size. The vertical axis shows the Lyapunov exponents for four simulations with $128$, $256$, $512$, and $1024$ cells in $\hat{z}$ and one in $\hat{x}$ and $\hat{y}$ directions. The domain size is $1\times 1\times 64$ cm, and there are $24,088$ particles in each cell. The number of cells in the simulation has a slight effect on the Lyapunov exponents. In this work, we consider convergence achieved for $1,024$ cells in a domain size of $1\times 1\times 64$ cm.}
            \label{fig: convergence_number_of_cells}
        \end{figure}
        
        To investigate the impact of the cells size on the Lyapunov exponents, we perform four simulations with $128$, $256$, $512$, and $1024$ cells in the $\hat{z}$ direction and one in the $\hat{x}$ and $\hat{y}$ directions. The domain size is $1\times 1\times 64$ cm, and there are $24,088$ particles per cell. We introduce a perturbation $| \vec{\delta}_{t_0} |/| \vec{r}_{t_0} |\sim 10^{-10}$ in a random direction at 2.65 ns. The simulation concludes at 20 nanoseconds or when $| \vec{\delta}_{t} |/| \vec{r}_{t} |\sim 10^{-6}$.
        
        Figure \ref{fig: convergence_number_of_cells} shows that the Lyapunov exponents are slightly affected by the cell size. We consider convergence achieved for $1,024$ cells within a domain size of $1\times 1\times 64$ cm.

\section{Flavor vector magnitude}
\label{appendixb}
    
    We aim to demonstrate that the magnitude of the flavor vector, as defined by Equation \eqref{eq: flavor_vector}, remains constant over time. If this statement is true, the trajectories of the flavor vector in the state space are bounded to the surface of a $16N_\mathrm{par}$-dimensional sphere of radius the length of the flavor vector. This proof simplifies to show that the flavor polarization vector $P_l^k$ maintains a constant magnitude, given that the number of neutrinos $N^k$ carried by each computational particle in the simulation remains unchanged throughout. To begin, we expand the density matrix and Hamiltonian as vectors of coefficients of the Gell-Mann matrices $G_i$:
    \begin{eqnarray}
        \rho &=& P_i G_i\label{rho}\\
        H &=& H_i G_i \label{H}\,\,,
    \end{eqnarray}
    where $\rho_i$ and $H_i$ are real numbers. Einstein summation convention is assumed for repeated indices. The time evolution of the density matrix is described by the following expression
    \begin{eqnarray}
        \frac{\partial \rho}{\partial t}&=&-i\left[H,\rho\right].
    \end{eqnarray}
    In terms of Gell-Mann vectors in \eqref{rho} and \eqref{H} the equation becomes
    \begin{eqnarray}
        \frac{\partial P_i}{\partial t} G_i &=&-iH_jP_k\left[G_j, G_k \right],
    \end{eqnarray}
    here $\left[G_j, G_k \right] = 2if^{jkl}G_l$, where the structure constants $f^{ijk}$ are completely antisymmetric in the three indices. 
    The time evolution of $P_i$ is now given by
    \begin{eqnarray}
        \frac{\partial P_i}{\partial t} G_i &=& 2 H_jP_k f^{jkl}G_l
    \end{eqnarray}
    This implies that
    \begin{eqnarray}
        \frac{\partial P_i}{\partial t} &=& 2 H_jP_k f^{jki}.
    \end{eqnarray}
    $\vec{P}\cdot \dot{\vec{P}}$ will be given by 
    \begin{eqnarray}
        P_i \frac{\partial P_i}{\partial t} &=& 2 H_jP_iP_k f^{jki} = 0, \\
    \end{eqnarray}
    since $H_jP_iP_k f^{jki} = H_jP_kP_i f^{jik} = -H_jP_iP_k f^{jki}$. This demonstrates that $P_i$ and the flavor vector in Equation \eqref{eq: flavor_vector} only change their orientation but not their magnitude.

\bibliographystyle{apsrev4-2}
\bibliography{refs}

\end{document}